\documentclass[journal]{IEEEtran}
\usepackage{cite}
\usepackage{amsmath}
\usepackage{graphicx}
\usepackage{tikz}
\usepackage{pgfplots}
\usepackage{mathtools}
\usepgfplotslibrary{colorbrewer}
\usepgfplotslibrary{groupplots}
\usepgfplotslibrary{fillbetween}
\usetikzlibrary{shapes.geometric}
\usetikzlibrary{arrows.meta}
\usepackage{pgfplotstable}
\pgfplotsset{compat=1.18}
\usetikzlibrary{backgrounds}
\usepackage{wrapfig}  
\usepackage{subfigure}
\usepackage{amsfonts}
\usepackage{orcidlink}

\newcommand{\ASE}{\ensuremath{\text{ASE}}}

\newcommand{\NLI}{\ensuremath{\text{NLI}}}
\newcommand{\SNR}{\ensuremath{\mathrm{SNR}}}

\DeclareMathOperator{\atan}{atan}
\DeclareMathOperator{\sign}{sign}

\DeclareMathOperator{\Si}{Si}

\usepgfplotslibrary{colorbrewer,fillbetween,groupplots,colormaps}
\usepackage{xcolor}
\definecolor{green1}{RGB}{27,158,119}
\definecolor{orange1}{RGB}{217,95,2}
\definecolor{purple1}{RGB}{117,112,179}
\definecolor{pink1}{RGB}{231,41,138}
\definecolor{gray1}{RGB}{102,102,102}

\definecolor{matlab1}{RGB}{0,114,189}
\definecolor{matlab2}{RGB}{217,83,25}
\definecolor{matlab3}{RGB}{237,177,32}
\definecolor{matlab4}{RGB}{126,47,142}
\definecolor{matlab5}{RGB}{119,172,48}
\definecolor{matlab6}{RGB}{77,190,238}
\definecolor{matlab7}{RGB}{162,20,47}

\begin{document}

\title{\huge A Closed-form Expression of the Gaussian Noise Model Supporting O-Band Transmission}

\author{
Zelin~Gan~\orcidlink{0000-0001-6911-264X},
Henrique~Buglia~\orcidlink{0000-0003-1634-0926}, %
Romulo~Aparecido, %
Mindaugas~Jarmolovi\v{c}ius~\orcidlink{0000-0002-0456-110X}, %
Eric~Sillekens~\orcidlink{0000-0003-1032-6760}, %
Jiaqian~Yang~\orcidlink{0000-0002-4564-6687}, %
Ronit~Sohanpal, %
Robert~I.~Killey %
 and Polina~Bayvel~\orcidlink{0000-0003-4880-3366} %
\thanks{This work was supported by EPSRC Programme Grant  TRANSNET (Transforming Networks - building an intelligent optical infrastructure EP/R035342/1) and EWOC (Extremely Wideband Optical Fibre Communication Systems,  EP/W015714/1). M.~Jarmolovi\v{c}ius is funded by an EPSRC studentship (EP/T517793/1), the Microsoft 'Optics for the Cloud' Alliance and an UCL Faculty of Engineering Sciences Studentship. Polina Bayvel is supported by a Royal Society Research Professorship.}  %
\thanks{Zelin~Gan, Romulo~Aparecido, Mindaugas~Jarmolovi\v{c}ius, Eric~Sillekens, Jiaqian~Yang, Ronit~Sohanpal, Robert~I.~Killey, and Polina~Bayvel are with the Optical Networks Group, Department of Electronic and Electrical Engineering, Torrington Place, University College London, WC1E 7JE, London, U.K. (e-mail: zelin.gan.17@ucl.ac.uk)} %
\thanks{Henrique~Buglia was with the Optical Networks Group and is now with Nokia, San Jose, CA, USA.}
}

\markboth{Journal of Lightwave Technology,~Vol.~XX, No.~XX, XX~2025}%
{Gan \MakeLowercase{\textit{et al.}}}

\maketitle

\begin{abstract}
We present a novel closed-form model for nonlinear interference (NLI) estimation in low-dispersion O-band transmission systems. The formulation incorporates the four-wave mixing (FWM) efficiency term as well as the coherent contributions of self- and cross-phase modulation (SPM/XPM) across multiple identical spans. This extension enables accurate evaluation of the NLI in scenarios where conventional closed-form Gaussian Noise (GN) models are limited. The proposed model is validated against split-step Fourier method (SSFM) simulations and numerical integration across 41-161 channels, with a 96~GBaud symbol rate, bandwidths of up to 16.1~THz, and transmission distances from 80 to 800~km. Results show a mean absolute error of the NLI signal-to-noise ratio (SNR) below 0.22~dB. The proposed closed-form model offers an efficient and accurate tool for system optimisation in O-band coherent transmission.
\end{abstract}


\IEEEpeerreviewmaketitle

\section{Introduction}
\label{Introduction}

\IEEEPARstart{T}{he} relentless growth in global data traffic has inspired research on optimising the use of the optical fibre bandwidth beyond the C-band. 

Much of the work has focused on transmission across S+C+L-bands~\cite{Jiaqian2025Transmission} and, more recently, into the U-, E-, and O-band, reaching bandwidths in excess of 37~THz~\cite{Puttnam2024OFC}. In contrast to the ultra-wideband (UWB) systems, which rely predominantly on different types of optical amplifiers, e.g., thulium- (T-) doped fibre amplifiers (DFAs), distributed Raman amplification, and are potentially limited by the use of band (de)multiplexers and inter-channel stimulated Raman scattering (ISRS), the band known as the O- (the 'original' band), benefits from the development of bismuth-doped fiber amplifiers (BDFAs); a single amplifier of this type covers the entire O-band from 1260 to 1360~nm. In this band, the low dispersion regime in standard single-mode fibers also minimises the need for chromatic dispersion compensation and thus simplifies digital signal processing (DSP) in the transceiver, for both intensity-modulated direct-detection (IM-DD) and coherent systems. This makes the O-band attractive for intra/inter data centre interconnects (DCIs)~\cite{Seiler2021Toward}. Recently, coherent O-band transmission has been demonstrated in~\cite{Daniel2024JLT,Daniel2025Continuous}, achieving over 106~Tb/s with a 16.4~THz bandwidth. However, the intrinsically higher nonlinear distortions in the O-band, due to near-zero dispersion and stronger phase matching, requires new approaches to efficient and accurate modelling of signal propagation in this regime.  These models are essential to understand the impact of nonlinearities on the quality of transmission and to optimise system performance

The quality of transmission (QoT) in an optical fibre system is often evaluated using the Gaussian Noise (GN) model. This provides an integral estimate of nonlinear interference (NLI) generated by Kerr effects~\cite{gnmodel}. Over the past ten years, it has been extended to include the term caused by the ISRS effect~\cite{isrsgnmodel,MattiaISRSGNmodel} in UWB systems from C- to C+L-band and beyond. %
Closed-form formulas for the models described in~\cite{isrsgnmodel,ISRSGNmodel_correction} were obtained by neglecting FWM, i.e., by considering only  SPM and XPM contributions to the NLI noise. This is because FWM is negligible for systems in which dispersion values are neither zero nor close to zero, which is the case for systems operating in the E-, S-, C-, L- and U-band over standard SMFs. Closed-form models (CFMs) using this approach were developed for lumped-amplified links~\cite{closed_gauss_daniel,ISRSGNmodel_correction}, and later updated to account for arbitrary span length and fibre losses~\cite{Buglia_JLT_lowloss}. Further CFMs, which do not yet account for FWM contributions, were developed for links using Raman, in combination with lumped amplification ~\cite{Buglia_JLT2024_CFRaman,Buglia_JLT2024_UWBmodelling,poggiolini2022closed}. 

To include the O-band in these models, where dispersion values are small or even zero, FWM contributions must be taken into account in the NLI noise estimation. Additionally, the low dispersion in the O band reduces the phase-mismatch term, meaning the NLI remains more coherent across spans in long-haul transmission. Consequently, the assumption in~\cite{danielclosed,Buglia_JLT2024_CFRaman,Buglia_JLT2024_UWBmodelling} that XPM accumulates incoherently does not hold in this regime. The inclusion of FWM contributions in the CFM in~\cite{Buglia_JLT_lowloss} was carried out in~\cite{Filippos2024Extended,Filippos2025Extending}, making it valid for O-band transmission systems. However, these O-band CFMs are limited to a single span, the model only includes up to the third-order GVD parameter ($\beta_3$), not all FWM contributions to the NLI are included, and fitting optimisation is needed for each FWM efficiency term as opposed to each channel, increasing model complexity.

In this work, we present a new closed-form formulation of the FWM contributions that addresses all the aforementioned limitations. The proposed model: 1) accounts for all FWM components contributing to NLI noise; 2) extends the phase mismatch term to include up to the fourth-order GVD parameter ($\beta_4$); 3) introduces new coherent contributions of SPM and XPM, which are essential for accurate NLI estimation in the O-band over multiple spans, ensuring the model remains valid for any number of spans; 4) requires the fitting optimization to be carried out for each channel, not increasing the model complexity, made possible by new closed-form approximations of the FWM efficiency term and normalised signal power profile evolution along the fibre length.

The proposed formula is valid for Gaussian constellations, and its accuracy is validated using integral model and SSMF simulations using 96~GBaud WDM channels in the O-band, covering 4.1 to 16.1~THz bandwidth, 1$\times$80~km to 10$\times$80~km distances, and a range of launch powers. The rest of the paper is organised as follows. In Section~\ref{sec:The ISRS GN model}, we formulate the ISRS GN model in integral form with the FWM contributions, and present the closed-form for single and multi-span transmissions. The application of the new closed-form model is demonstrated in Section~\ref{sec:Results} over multiple transmission scenarios, and its accuracy is validated. Key results are summarized in Section~\ref{sec:Conclusion}. Appendix~\ref{appA:link_function} describes the derivation of the link function, and Appendix~\ref {appB:FWM} contains the derivation of the expressions for the closed-form FWM coefficient. The self-phase modulation (SPM) and cross-phase modulation (XPM) coherent contributions in multi-span transmission are derived in Appendices~\ref{appC:SPM_multispan}~and~\ref{appD:XPM_multispan}, respectively. All mathematical identities used in this paper are listed in Appendix~\ref{appE:mathematical_identities}.

\section{The ISRS GN model and inclusion of FWM contributions}
\label{sec:The ISRS GN model}
This section describes the ISRS GN model used to semi-analytically estimate the NLI noise. The inclusion of FWM in the integral model described in \cite{danielclosed,Buglia_JLT_lowloss} is presented together with its closed-form approximation. New coherence factor derivations for SPM and XPM are also presented. These new contributions are used in conjunction with the model in~\cite{Buglia_JLT_lowloss} to estimate the system performance. The mathematical derivation of these formulas is described in the Appendices. 

For an ideal transceiver, after coherent detection and electronic dispersion compensation, the total received $\SNR$ for the $i$-th WDM channel ($\SNR_{i}$) after $N_{\rm s}$ spans can be estimated as
\begin{equation}
\begin{aligned}
&\SNR_{i} \approx\frac{P_i}{N_{\rm s} P_{\ASE_{i}} + \eta_{\text{NLI}}(f_i)P_i^3},
\end{aligned}
\label{eq:snr}
\end{equation}
where $\SNR_{\text{ASE}_i} = \frac{P_i}{N_{\rm s}P_{\text{ASE}_i}}$ and $\SNR_{\text{NLI}_i} = \frac{P_i}{\eta_{\text{NLI}}(f_i)P_i^3}$ are the $\SNR$ values due to the amplified spontaneous emission (ASE) from the optical amplifiers used to compensate for the fibre loss, and the accumulated $\NLI$, respectively. $N_{\rm s}$ is the number of spans, $i$ is the channel of interest (COI), $P_i$ is its launch power, $P_{\ASE_{i}}$ is the $\ASE$ noise power at the $i$-th channel frequency, and $P_{\NLI_{i}} = \eta_{\text{NLI}}(f_i)P_i^3$ is the $\NLI$ noise power after $N_{\rm s}$ spans. This paper focuses on the $\SNR_{\text{NLI}}$ calculation.

To calculate the power spectral density~(PSD) of the $i$-th channel $P_{\text{NLI}_i}$, the ISRS GN model approach is considered. The inclusion of FWM in the model published in~\cite{Buglia_JLT_lowloss} is presented in this section, enabling an accurate performance estimation of lumped-amplified systems operating over the O-to-U-band, or using the O-band alone.

\subsection{The ISRS GN Model in Integral Form}
\label{The ISRS GN Model in Integral Form}

\begin{figure*}
\centering
\begin{tikzpicture}
\begin{groupplot}[
  group style={group size=2 by 1, horizontal sep=1.2cm-6.8583pt},
  width=0.56\textwidth,        %
  height=0.56\textwidth,
  axis equal image,
  axis lines=none, ticks=none, %
  clip=false
]

\nextgroupplot
\addplot graphics[xmin=0,xmax=1,ymin=0,ymax=1] {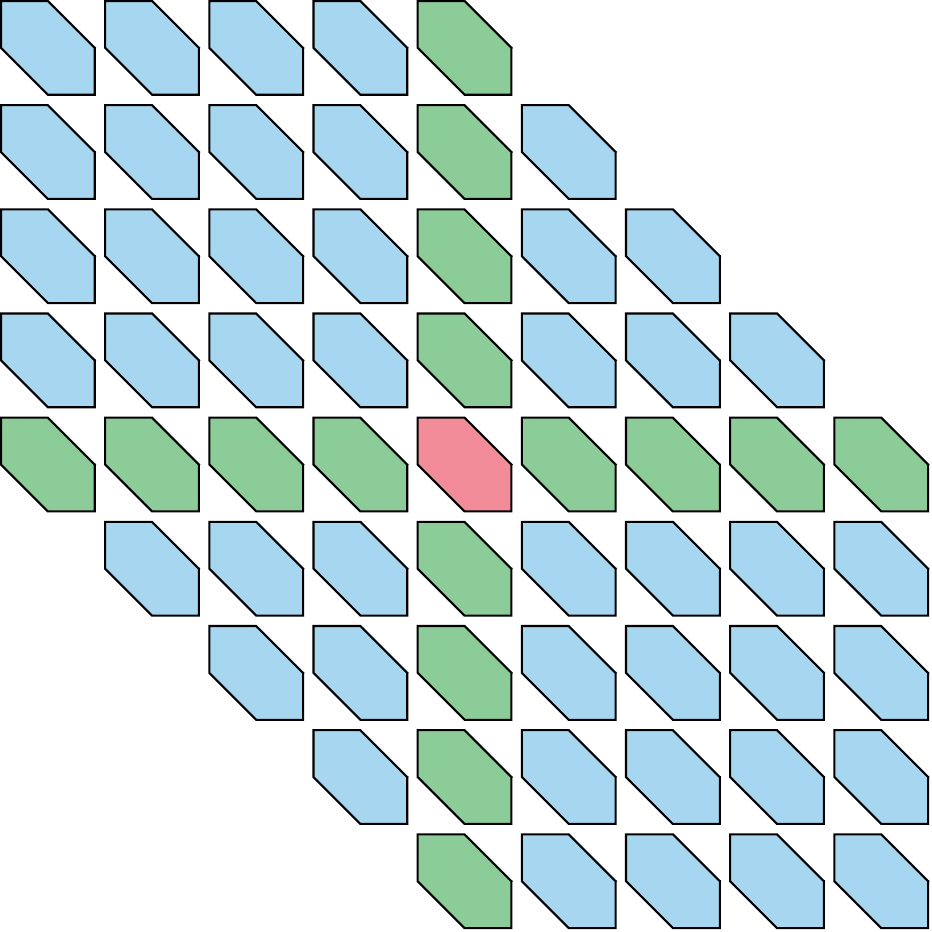};

\draw[-{Stealth[length=2.2mm]}, thick]
  (rel axis cs:0,0.5) -- (rel axis cs:1,0.5);   %
\draw[-{Stealth[length=2.2mm]}, thick]
  (rel axis cs:0.5,0) -- (rel axis cs:0.5,1);   %
\draw[-{Stealth[length=2.2mm]}, thick]
  (rel axis cs:0.25,0.25) -- (rel axis cs:0.94,0.94);   %

\node[anchor=south] at (rel axis cs:0.5,1.02) {(a) Channel $i=5$};
\node[anchor=north east] at (rel axis cs:1.00,0.5) {$f_1$};
\node[anchor=north west] at (rel axis cs:0.5,1.00) {$f_2$};
\node[anchor=north east] at (rel axis cs:1.00,1.00) {$f_3 = f_1 + f_2 - f_i$};

\nextgroupplot
\addplot graphics[xmin=0,xmax=1,ymin=0,ymax=1] {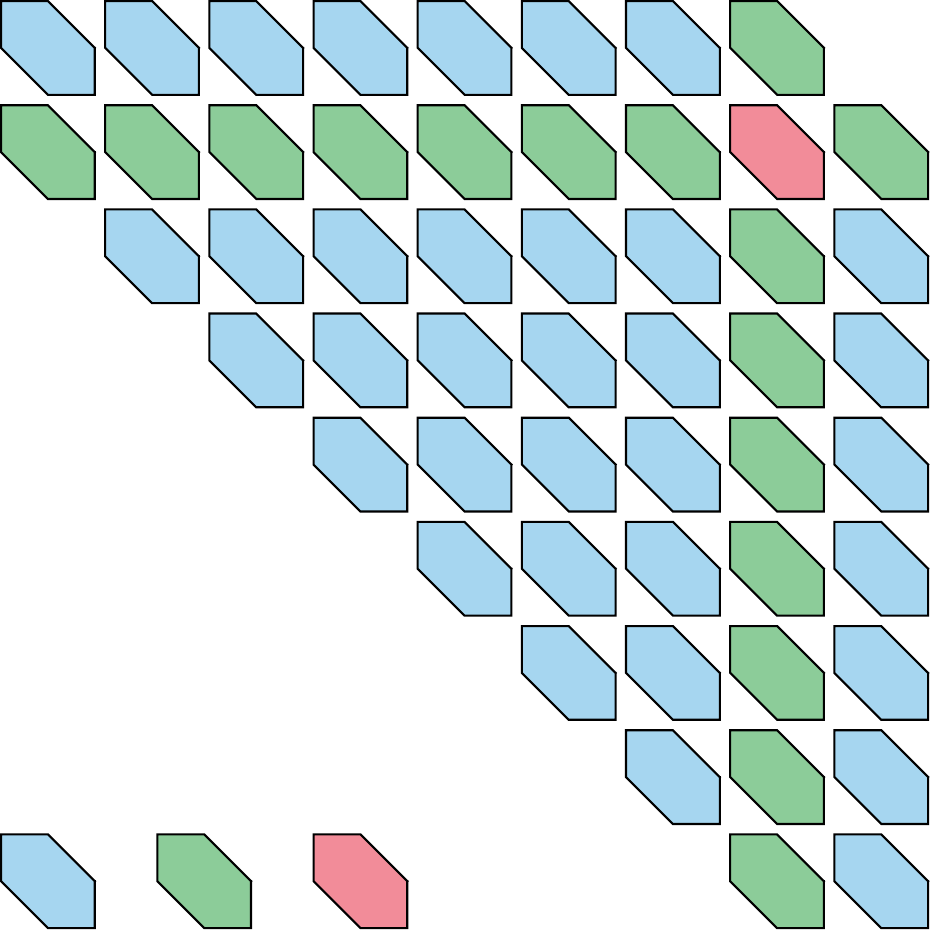};

\draw[-{Stealth[length=2.2mm]}, thick]
  (rel axis cs:0,0.5) -- (rel axis cs:1,0.5);   %
\draw[-{Stealth[length=2.2mm]}, thick]
  (rel axis cs:0.5,0) -- (rel axis cs:0.5,1);   %
\draw[-{Stealth[length=2.2mm]}, thick]
  (rel axis cs:0.25,0.25) -- (rel axis cs:0.94,0.94);   %

\node[anchor=south] at (rel axis cs:0.5,1.02) {(b) Channel $i=8$};
\node[anchor=north east] at (rel axis cs:1.00,0.5) {$f_1$};
\node[anchor=north west] at (rel axis cs:0.5,1.00) {$f_2$};
\node[anchor=north east] at (rel axis cs:1.00,1.00) {$f_3 = f_1 + f_2 - f_i$};

\node[anchor=south west] at (rel axis cs:0.05,0.17) {FWM};
\node[anchor=south west] at (rel axis cs:0.19,0.17) {XPM};
\node[anchor=south west] at (rel axis cs:0.33,0.17) {SPM};

\end{groupplot}
\end{tikzpicture}
\caption{Integration domain for NLI calculation of a 9-channel WDM system at (a) channel $i=5$ (centre channel) and (b) channel $i=8$, where SPM, XPM, and FWM contributions are in red, green, and blue respectively.}
\label{fig:integration_domain}
\end{figure*}
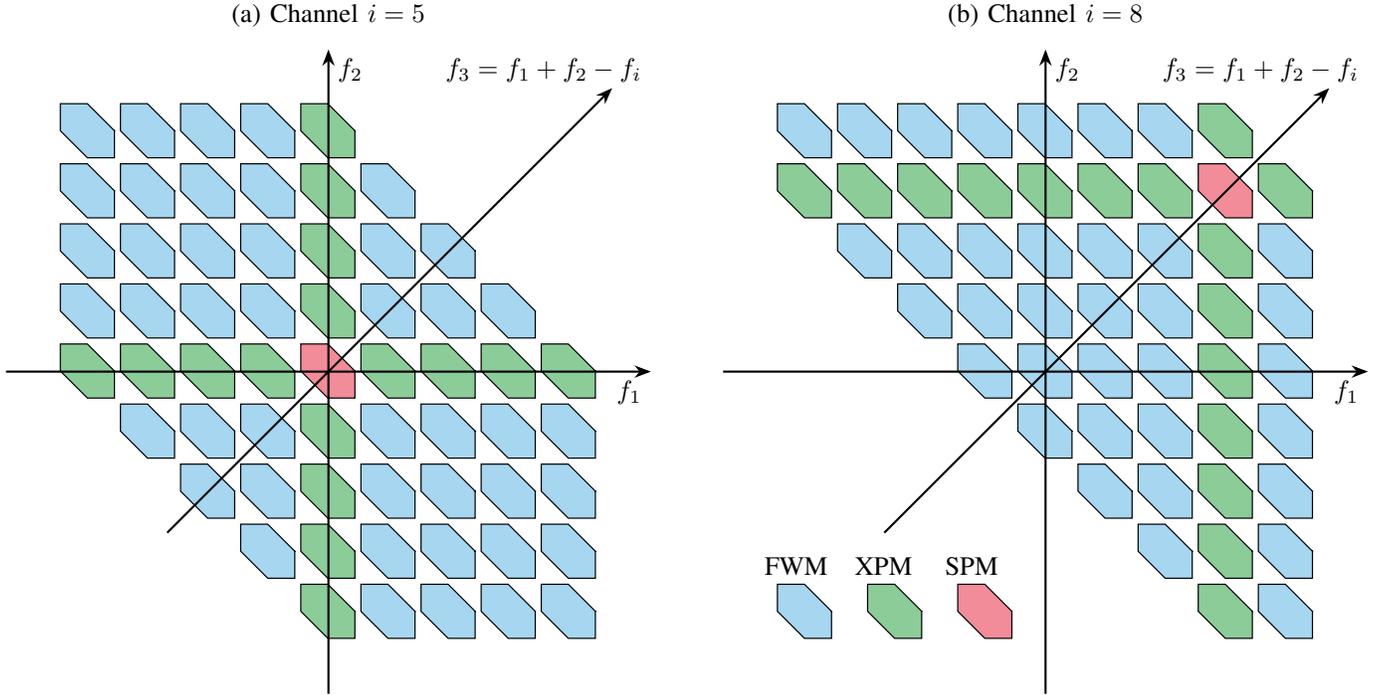 
The nonlinear coefficient contribution accounting for Gaussian modulated symbols can be written as
\begin{equation}
\begin{split}
&\eta_{\text{NLI}}(f_i) = \frac{16}{27} \gamma^2 \frac{ B_i}{P_i^3} \int df_1 \int df_2 G(f_1)G(f_2)G(f_1 + f_2 - f_i) \\
&\times \mu(f_1,f_2,f_i) \  \chi\left(f_1, f_2, f_i\right),
\label{eq:GN_integral_gaussian}
\end{split}
\end{equation}
where $\gamma$ is the optical fibre nonlinearity coefficient, $B_i$ is the bandwidth of the COI, $G(\cdot)$ represents power spectral density, and $\mu(f_1, f_2, f_i)$ is the so-called link function or FWM efficiency given by
\begin{equation}
\begin{split}
&\mu\left(f_1,f_2,f_i\right)= \\
&\left| \int_0^L d\zeta \ \sqrt{\frac{\rho(\zeta,f_1) \rho(\zeta,f_2) \rho(\zeta,f_1 + f_2 - f_i)}{\rho(\zeta,f_i)}} e^{j\phi\left(f_1,f_2,f_i\right)\zeta}\right|^2,
\label{eq:link_function_integral}
\end{split}
\end{equation}
where $\rho$ is the normalised signal power profile and $\phi$ the phase-mismatch term defined as
\begin{equation}
\begin{split}
&\phi(f_1,f_2,f_i) = -4\pi^2\left(f_1-f_i\right)\left(f_2-f_i\right)\Big[\beta_2+\pi\beta_3\left(f_1+f_2\right) \\
&+ \frac{2\pi^2}{3} \beta_4[(f_1 - f_i)^2 + \frac{3}{2}(f_1-f_i)(f_2-f_i) + 3(f_1 - f_i)f_i \\
&+ (f_2-f_i)^2 + 3(f_2-f_i)f_i + 3f_i^2]\Big].
\label{eq:mismatch_term}
\end{split}
\end{equation}
$\beta_2$, $\beta_3$ and $\beta_4$ are the second, third and fourth-order GVD parameters, respectively~\cite{Min_JLT}. The phased-array term $\chi\left(f_1, f_2, f_i\right)$ accounts for the coherent interference of the NLI in each span of multi-span systems, where each span has identical fibre parameters and signal power profiles. It has the form
\begin{equation}
\chi\left(f_1, f_2, f_i\right) = \left|\frac{\sin\left(\frac{1}{2}N_{\rm s}\phi\left(f_1, f_2, f_i\right)L\right)}{\sin\left(\frac{1}{2}\phi\left(f_1, f_2, f_i\right)L\right)}\right|^2,
\label{eq:phased_array_factor}
\end{equation}
where $L$ is the span length.

Eq.~\eqref{eq:GN_integral_gaussian} considers all the nonlinear contributions (SPM, XPM and FWM) to the total NLI noise. To derive closed-form expressions, it is convenient to write these contributions separately. %
Thus, $\eta_{\text{NLI}}(f_i)$ can be written as
\begin{equation}
\begin{split}
&\eta_{\text{NLI}}(f_i) = \eta_{\text{SPM}}(f_i) + \eta_{\text{XPM}}(f_i) + \eta_{\text{FWM}}(f_i),
\label{eq:eta_separate_gauss}
\end{split}
\end{equation}
where $\eta_{\text{SPM}}(f_i)$ is the SPM contribution, $\eta_{\text{XPM}}(f_i)$ is the total XPM contribution, and $\eta_{\text{FWM}}(f_i)$ is the total FWM contribution to the NLI, all generated after $N_{\rm s}$ spans.

The closed-form expressions for the first-span SPM and XPM contributions in Eq.~\eqref{eq:eta_separate_gauss}, $\eta_{\text{SPM}}(f_i)$ and $\eta_{\text{XPM}}(f_i)$, were previously obtained~\cite{closed_gauss_daniel,Buglia_JLT_lowloss}. It was assumed that only SPM accumulates coherently, through a coherent factor~\cite{gnmodel}. In this work, we separate incoherent and coherent contributions of both SPM and XPM - this is essential for accurate estimation of NLI in O-band; further details are provided in Section~\ref{sec:Multi-span}. To account for different fibre parameters and launch power in each span, the FWM contribution is assumed to accumulate incoherently and is given by
\begin{equation}
\begin{split}
&\eta_{\text{FMW}}(f_i)  =\sum_{q=1}^{N_{\rm s}} \left( \frac{P_{i,q}}{P_i} \right)^2 \sum_{\substack{j,k,m \in \Omega}}^{N_{\text{ch}}} \eta_{\text{FMW},q}^{(j,k,m)}(f_i),
\label{eq:FWM_gauss}
\end{split}
\end{equation}
where $P_{i,q}$ is the power of channel $i$ launched into the $q$-th span, $N_{\text{ch}}$ is the number of WDM channels and $\eta_{\text{FMW},q}^{(j,k,m)}(f_i)$ is the FMW contribution generated in the $q$-th span of interfering channels $j$, $k$, and $m$, with frequencies $f_j$, $f_k$ and $f_m$ on channel $i$ with frequency $f_i$. For simplicity, identical spans are considered for the remainder of this work and the $q$ dependence of the FWM contribution is suppressed below. 

Not all the combinations of the channels with frequencies $f_j$, $f_k$ and $f_{m}$ generate a nonlinear distortion in the channel with frequency $f_i$. The combinations that contribute to the nonlinear distortion in the channel $f_i$ are those satisfying
\begin{equation}
f_i = f_j + f_k - f_m.
\label{eq:FWM_condition}
\end{equation}
These contributions can be classified in SPM and XPM if $f_j = f_i$ and $f_k = f_m$, or $f_k = f_i$ and $f_j = f_m$. The remaining interactions satisfying Eq.~\eqref{eq:FWM_condition}, which are not SPM and XPM, are called FWM contributions. Thus, a set representing all related frequency triples $\left(f_j, f_k, f_m\right)$ can be defined as $\Omega$
\begin{equation}
\begin{split}
&\Omega = \{\left(f_j, f_k, f_m\right) \: | \: f_j + f_k - f_m = f_i, \\ 
&f_j \neq f_i \: \&  \: f_k \neq f_m, f_k \neq f_i \: \& \: f_j \neq f_m\}.
\label{eq:omega_region}
\end{split}
\end{equation}
An example of the integration domain of a 9-channel WDM system is shown in Fig.~\ref{fig:integration_domain}. The FWM contributions are represented by blue regions. The FWM contributions of a single interfering channels after a single span with frequencies $f_j$, $f_k$ and $f_{m}$ can be written as
\begin{equation}
\begin{split}
&\eta_{\text{FWM}}^{(j,k,m)}(f_i) = \frac{16}{27} \gamma^2 \frac{B_i}{P_i^3} \frac{P_j P_k P_m}{B_j B_k B_m} \int_{\frac{-B_j}{2}}^{\frac{B_j}{2}} df_1 \int_{\frac{-B_k}{2}}^{\frac{B_k}{2}} df_2 \ \\ 
&\times \Pi \left(\frac{f_1 + f_2}{B_m} \right) \left| \mu (f_1+f_j, f_2 + f_k, f_i) \right|^2,
\label{eq:FWM_GN_integral}
\end{split}
\end{equation}
where $P_j$, $B_j$, $P_k$, $B_k$, $P_m$ and $B_m$ are the power and bandwidth of channels $j$, $k$ and $m$, and $\Pi(x)$ denotes the rectangular function.

\subsection{The Closed-form Expression}
\label{sec:closedform_expression}
This section presents an analytical expression of Eq.~\eqref{eq:FWM_GN_integral}. Note that, this equation depends on the link function, given by Eq.~\eqref{eq:link_function_integral}. Thus, the first step is to derive a closed-form expression of Eq.~\eqref{eq:link_function_integral}, and then use it to obtain closed-form expression of Eqs.~\eqref{eq:FWM_GN_integral}. 

 In the case of FWM, Eq.~\eqref{eq:link_function_integral} involves at least three different channels in the set $\Omega$ and satisfying Eq.~\eqref{eq:omega_region}. The interactions involving four different channels correspond to the frequencies $f_j$, $f_k$, $f_m$ and $f_i$. The interactions between three different channels correspond to the frequencies $f_j$, $f_k$ and $f_m = f_i$, or $f_j = f_k$, $f_m$ and $f_i$. Let $x = P_{\text{tot}} C_{r,i} L_{\text{eff}}(z)$. The normalised signal profile evolution along the
fibre length $z$ for an arbitrary channel $i$ can be approximated with Eq.~(17) in~\cite{closed_gauss_daniel}
\begin{equation}
\begin{split}
&\rho(z,f_i) = \frac{P(z,f_i)}{P(0,f_i)} \approx \frac{B_{\text{tot}}x e^{-x f_i}}{2 \sinh{\left(\frac{B_{\text{tot}}}{2}x\right)}}e^{-\alpha_i z}, 
\label{eq:signal_profile_closed}
\end{split}
\end{equation}
where $B_{\text{tot}}$ is the total bandwidth of WDM spectra, $L_{\text{eff}}(z) = \frac{1-e^{-\tilde{\alpha}_i z}}{\tilde{\alpha}_i}$, $P_{\text{tot}}$ is the total launch power, $\alpha_i$ and $\tilde{\alpha}_i$ model the fibre loss, and $C_{r,i}$ is the slope of the Raman gain spectrum. 
Similar to Eq.~(13) in~\cite{Buglia_JLT_lowloss}, the square root of the normalised signal profile evolution along the
fibre distance for an arbitrary channel $i$, given by Eq.~\eqref{eq:signal_profile_closed}, can be obtained approximately by using a first-order Taylor expansion around $x = 0$ for the square root of the fraction in the right-hand side of Eq.~\eqref{eq:signal_profile_closed}, yielding:
\begin{equation}
\begin{split}
&\sqrt{\rho(z,f_i)} \approx e^{-\frac{\alpha_i}{2} z}\left(1 - \frac{P_{\text{tot}}C_{r,i} f_i L_{\text{eff}}(z)}{2}\right), 
\label{eq:signal_profile_closed_taylor}
\end{split}
\end{equation}
where the coefficients $\alpha_i$, $\tilde{\alpha}_i$ and $C_{r,i}$ are calculated using the fitting strategy based on Eq.~(13) in~\cite{Buglia_JLT_lowloss}. Eq.~\eqref{eq:signal_profile_closed_taylor} can be simply written for other channels by replacing the index $i$ by $k$, $j$ or $m$. The fitting is calculated for each WDM channel, and these pre-computed values are inserted in the formulas obtained in this section to obtain the total NLI noise for each WDM channel. Thus, the total number of fittings is the same as the number of channels $N_{\text{ch}}$.

Let $\tilde{T}_i = -\frac{P_{\text{tot}}C_{r,i}}{2\tilde{\alpha}}f_i$, $T_i = 1 + \tilde{T}_i$. By assuming the square root of the normalised power evolution along the fibre $\sqrt{\rho(z,f_i)}$ as Eq.~\eqref{eq:signal_profile_closed_taylor}, the link function in Eq.~\eqref{eq:link_function_integral} can be obtained in closed form for two different subsets of frequency combinations in the set $\Omega$, where $\Omega_{1}+\Omega_{2} = \Omega$. For the three different frequency combinations $f_j$, $f_k$ and $f_m$, the subset $\Omega_{1} = \{\left(f_j, f_k, f_m\right)\in \Omega \: | \: f_m=f_i\}$. For the remaining frequencies in $\Omega$, the subset $\Omega_{2} = \{\left(f_j, f_k, f_m\right)\in \Omega \: | \: f_m\neq f_i\}$, and $P_{\text{tot}}C_{r,i} f_i L_{\text{eff}}(z) \approx 0$ is assumed in Eq.~\eqref{eq:signal_profile_closed_taylor} for the channel $i$ only. This is the same as neglecting part of the ISRS effect for that channel, as part of this effect is also captured by $\alpha_i$ after fitting optimisation. To distinguish the link functions in $\Omega_{1}$ and $\Omega_{2}$ while retaining a single compact expression, we define
\begin{equation}
\ell =
    \begin{cases}
      \left(l_j, l_k, l_m\right) \in \{0,1\}^3 & \text{if $\left(f_j, f_k, f_m\right)\in \Omega_1$}, \\
      \left(l_j, l_k\right) \in \{0,1\}^2 & \text{if $\left(f_j, f_k, f_m\right)\in \Omega_2$},
      \label{eq:link_simplify_l}
    \end{cases} 
\end{equation}
and a variable
\begin{equation}
\delta_{\Omega} =
    \begin{cases}
      1 & \text{if $\left(f_j, f_k, f_m\right)\in \Omega_1$}, \\
      0 & \text{if $\left(f_j, f_k, f_m\right)\in \Omega_2$}.
      \label{eq:link_simplify_d}
    \end{cases} 
\end{equation}
Thus, a closed-form expression of the link function for $\Omega_1$ and $\Omega_2$ is given by
\begin{equation}
\begin{split}
&\mu\left(f_1+f_j, f_2 + f_k, f_i\right) = \\
&\sum_{\substack{\ell, \ell'}} \mathcal{T}_{\ell}\mathcal{T}_{\ell'} \kappa_{\ell, i}\kappa_{\ell', i} \frac{\left(\tilde{\alpha}_{\ell, i}\tilde{\alpha}_{\ell', i}+\phi^2\right)}{\left(\tilde{\alpha}_{\ell, i}^2+\phi^2\right)\left(\tilde{\alpha}_{\ell', i}^2+\phi^2\right)},
\label{eq:link_function_closed_set3}
\end{split}
\end{equation}
where
\begin{equation}
\tilde{\alpha}_{\ell, i} = \frac{\alpha_{\ell, i}\left(1 - e^{-\alpha_{\ell, i}L}\right)}{1 - e^{-\alpha_{\ell, i}L} - \alpha_{\ell, i}e^{-\alpha_{\ell, i}L}},
\label{eq:alpha_tilde}
\end{equation}
and
\begin{equation}
\kappa_{\ell, i} = \frac{\tilde{\alpha}_{\ell, i}\left(1 - e^{-\alpha_{\ell, i}L}\right)}{\alpha_{\ell, i}}.
\label{eq:kappa}
\end{equation}
The variables $\alpha_{\ell, i}$ and $\mathcal{T}_{\ell}$ are given as
\begin{equation}
\alpha_{\ell, i} = l_j\tilde{\alpha}_j + l_k\tilde{\alpha}_k + \delta_{\Omega}l_m\tilde{\alpha}_m + \frac{\alpha_j + \alpha_k + \delta_{\Omega}\left(\alpha_m - \alpha_i\right)}{2},
\label{eq:link_simplify_a}
\end{equation}
and
\begin{equation}
\mathcal{T}_{\ell} = T_j T_k T_m^{\delta_{\Omega}} \left( \frac{-\tilde{T}_j}{T_j} \right)^{l_j} \left( \frac{-\tilde{T}_k}{T_k} \right)^{l_k} \left( \frac{-\tilde{T}_m}{T_m} \right)^{\delta_{\Omega}l_m}.
\label{eq:link_simplify_T}
\end{equation}
The proof of these equations is given in Appendix~\ref{appA:link_function}.

The next step is to use Eq.~\eqref{eq:link_function_closed_set3} to obtain a closed-form expression of Eq.~\eqref{eq:FWM_GN_integral}. A closed-form expression of $\eta_{\text{FWM}}^{(j,k,m)}(f_i)$ given in Eq.~\eqref{eq:FWM_GN_integral}, is given by
\begin{equation}
\begin{split}
&\eta_{\text{FWM}}^{(j,k,m)}(f_i) =\frac{16}{27} \tau \gamma^2 \frac{ B_i}{P_i^3} \\ 
&\times \frac{P_j P_k P_m}{\max{(B_j,B_k,B_m)}} \sum_{\substack{\ell, \ell'}} \frac{\mathcal{T}_{\ell}\mathcal{T}_{\ell'}\kappa_{\ell, i}\kappa_{\ell', i}}{\phi_1\phi_2\left(\tilde{\alpha}_{\ell, i}+\tilde{\alpha}_{\ell', i}\right)} \\
&\times \left[ \tilde{\alpha}_{\ell, i}\left(F\left(u_{+}\right)-F\left(u_{-}\right)-F\left(u'_{+}\right)+F\left(u'_{-}\right)\right) \right. \\
&\left. + \tilde{\alpha}_{\ell', i}\left(F\left(v_{+}\right)-F\left(v_{-}\right)-F\left(v'_{+}\right)+F\left(v'_{-}\right)\right) \right],
\label{eq:FWM_GN_closed}
\end{split}
\end{equation}
where $F\left(x\right) = x\atan\left(x\right)-\frac{1}{2}\ln\left(1+x^2\right)$ and 
\begin{equation}
\begin{aligned} 
u_{\pm} &= \frac{2\phi_0+\phi_1 B_j\pm\phi_2 B_k}{2\tilde{\alpha}_{\ell, i}},    & v_{\pm} &= \frac{2\phi_0-\phi_1 B_j\pm\phi_2 B_k}{2\tilde{\alpha}_{\ell, i}}, \\ 
u'_{\pm} &= \frac{2\phi_0+\phi_1 B_j\pm\phi_2 B_k}{2\tilde{\alpha}_{\ell', i}}, & v'_{\pm} &= \frac{2\phi_0-\phi_1 B_j\pm\phi_2 B_k}{2\tilde{\alpha}_{\ell', i}},
\end{aligned}
\end{equation}
and $\max{(B_j,B_k,B_m)}$ is the function which returns the maximum element between $B_j$, $B_k$ and $B_m$. $\phi_0$, $\phi_1$ and $\phi_2$ are coefficients of a first-order two-dimensional Taylor expansion of the phase-mismatch $\phi$ around $\left(f_1,f_2\right) = \left(0,0\right)$ where the expression is shown in Eqs.~\eqref{appB:mismatch_term_approx2}~and~\eqref{appB:mismatch_term_approx3}. The variable $\tau = 1$ for the three different frequency combinations $f_j = f_k$, $f_m$, and $f_i$. For all the remaining frequencies in $\Omega$, symmetry properties can be explored leading to $\tau = 2$.  The variables $\mathcal{T}_{\ell'}$, $\kappa_{\ell', i}$, $\tilde{\alpha}_{\ell', i}$, $\alpha_{\ell', i}$, $u'_{\pm}$, and $v'_{\pm}$ are the same as $\mathcal{T}_{\ell}$, $\kappa_{\ell, i}$, $\tilde{\alpha}_{\ell, i}$, $\alpha_{\ell, i}$, $u_{\pm}$, and $v_{\pm}$, but with the
indices $l_j$, $l_k$ and $l_m$ replaced by $l'_j$, $l'_k$ and $l'_m$. The use of the $\max$ function, along with the omission of $\Pi$ in Eq.~\eqref{eq:FWM_GN_integral}, ensures that the integration domain is approximated by a circumscribed rectangle in each region shown in Fig.~\ref{fig:integration_domain} in case of channels having different symbol rates. The proof of Eq.~\eqref{eq:FWM_GN_closed} is given in Appendix~\ref{appB:FWM}.

\subsection{Multi-span SPM and XPM Coherent Contribution}
\label{sec:Multi-span}
Many experiments have shown that, for multiple identical spans, the accumulation of NLI follows
\begin{equation}
P_{\text{NLI}}(f_i) = P_{\rm NLI,0}(f_i) N_{\rm s}^{1+\epsilon(f_i)},
\label{eq:epsilon}
\end{equation}
where $P_{\rm NLI,0}(f_i)$ is the $i$-th channel NLI power at the first span and $\epsilon$ quantifies the coherence with which NLI generated in different spans accumulates. $\epsilon = 0$ means that the NLI produced in all spans adds incoherently. Instead, the closer $\epsilon$ is to 1, the higher the coherence of the NLI contributions between different spans, with $\epsilon = 1$ corresponding to perfect phase-matching.

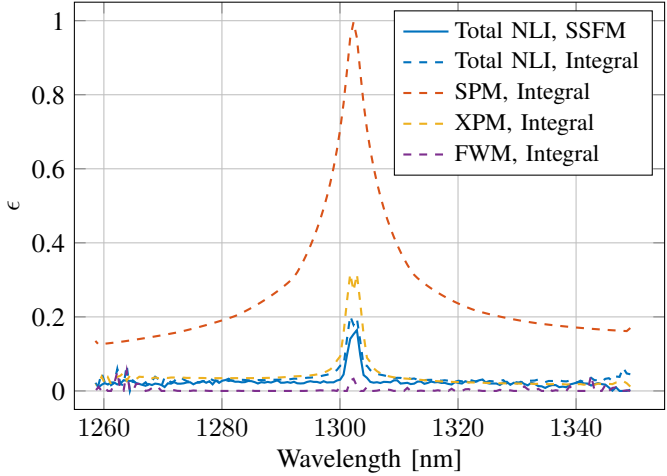
\begin{figure}

\begin{tikzpicture}
    \begin{axis}
    [
    legend columns=1,
    legend cell align=left,
    width=0.53\textwidth-5.21645pt,height=7cm,
    grid=both,
    legend style={fill opacity=1, draw opacity=1, text opacity=1, at={(0.98,0.98)}, anchor=north east, draw=black, font=\small},
    xlabel={Wavelength~[nm]},
    ylabel={$\epsilon$},
    ylabel near ticks,
    xlabel near ticks,
    ylabel shift = -2 pt,
    xlabel shift = -2 pt,
    ytick distance=1,
    ymin=-0.05,
    ymax=1.05,
    ytick={0,0.2,0.4,0.6,0.8,1},
    yticklabels={0,0.2,0.4,0.6,0.8,1},
    xmin=1255,
    xmax=1355,
    clip marker paths=true,
    tick label style={
        /pgf/number format/.cd,
        fixed,
        precision=3,
        use comma=false,
        1000 sep={},
        set decimal separator={.}
    },
    scaled ticks=false,
    axis x line*=box,
    ]

        \addplot[matlab1,thick,solid,mark=none,forget plot] table[x=wvec,y=eps_NLI_ssfm] {Data/eps_10span.tsv};  
        \addplot[matlab1,thick,dashed,mark=none,forget plot] table[x=wvec,y=eps_NLI_integral] {Data/eps_10span.tsv};
        
        \addplot[matlab2,thick,dashed,mark=none,forget plot] table[x=wvec,y=eps_SPM_integral] {Data/eps_10span.tsv};
        \addplot[matlab3,thick,dashed,mark=none,forget plot] table[x=wvec,y=eps_XPM_integral] {Data/eps_10span.tsv};
        \addplot[matlab4,thick,dashed,mark=none,forget plot] table[x=wvec,y=eps_FWM_integral] {Data/eps_10span.tsv};

        \addlegendimage{matlab1,thick,solid};
        \addlegendentry{Total NLI, SSFM};
        \addlegendimage{matlab1,thick,dashed};
        \addlegendentry{Total NLI, Integral};
        \addlegendimage{matlab2,thick,dashed};
        \addlegendentry{SPM, Integral};
        \addlegendimage{matlab3,thick,dashed};
        \addlegendentry{XPM, Integral};
        \addlegendimage{matlab4,thick,dashed};
        \addlegendentry{FWM, Integral};
         
    \end{axis}

\end{tikzpicture}

\caption{Coherence factor $\epsilon$ from a 10-span transmission simulation (solid line) and from the different contributions of the integral GN model (dashed line) using Eq.~\eqref{eq:epsilon}.}
\label{fig:eps_10span}

\end{figure}

For systems with more than one identical span $(N_{\rm s} > 1)$, the phased-array factor $\chi$ in Eq.~\eqref{eq:phased_array_factor} associated with the nonlinear coefficient becomes relevant, as shown in Eq.~\eqref{eq:GN_integral_gaussian}. The phase-array can be written in sum form as
\begin{equation}
\chi\left(f_1, f_2, f_i\right) = N_{\rm s} + 2\sum_{n=1}^{N_{\rm s}-1} \left(N_{\rm s}-n\right) \cos\left(n\phi\left(f_1, f_2, f_i\right)L\right).
\label{eq:phased_array_factor_sum}
\end{equation}
Then, the nonlinear coefficient can be split into incoherent and coherent contribution as follows
\begin{equation}
\eta_{\text{NLI}}(f_i) = \eta_{\text{NLI},\text{inc}}(f_i) + \eta_{\text{NLI},\text{cc}}(f_i),
\end{equation}
and it can be related to $\epsilon$ as shown in~\cite{gnmodel}:
\begin{equation}
\epsilon(f_i) = \log \left(1+\frac{\eta_{\text{NLI},\text{cc}}(f_i)}{\eta_{\text{NLI},\text{inc}}(f_i)}\right) \frac{1}{\log \left(N_{\rm s}\right)}.
\end{equation}
The coherent factor can be separated into three different contributions: SPM, XPM and FWM. The impact of the coherence among NLI in the O band is illustrated in Fig.~\ref{fig:eps_10span} of 16.1~THz bandwidth centred at 1302.3~nm 10-span transmission simulation. The coherence factors for total NLI and for its SPM, XPM, and FWM components all peak near the zero-dispersion wavelength. At this point SPM is perfectly phase-matched and coherent accumulation of NLI from this effect is the largest of all the contributions. XPM reaches a maximum $\epsilon$ around 0.32 which is non-negligible compared to a 10~THz bandwidth system centred at 1550~nm which has a $\epsilon$ of 0.15 on SPM and close to 0 on XPM~\cite{closed_gauss_daniel}. For FWM, the value of $\epsilon$ remains below 0.03, allowing it to be treated as incoherently accumulated.

The incoherent contribution of SPM and XPM in an identical multi-span system with ideal amplification at the end of each span can be obtained by simply multiplying to its single span SPM and XPM NLI contributions by a factor of $N_{\rm s}$ , but coherent contributions requires re-deriving the integral in~\cite[Eqs.~(5)]{Buglia_JLT_lowloss} including the sum form of the phase-array factor in Eq.~\eqref{eq:phased_array_factor_sum}. Let $\tilde{T}'_i = -\frac{P_{\text{tot}}C_{r,i}}{\tilde{\alpha}}f_i$, $T'_i = 1 + \tilde{T}'_i$. The coherent contribution term for SPM and XPM are given by
\begin{equation}
\begin{split}
&\eta_{\rm SPM,cc}(f_i) = \frac{16}{27}\frac{\gamma^2}{B_i^2} {T'_i}^2 \sum_{\substack{l, l' \in \{0, 1\}}} \left( \frac{-\tilde{T}'_i}{T'_i} \right)^{l+l'} \\
&\times \frac{\kappa_{l,i} \kappa_{l',i}}{\phi_i L\tilde{\alpha}_{l,i}\tilde{\alpha}_{l',i}} \sum_{n=1}^{N_{\rm s} - 1} \frac{8\left(N_{\rm s} - n\right)}{n} \atan\left(n\phi_i L\frac{B_i^2}{4}\right)
\label{eq:SPM_GN_closed_multispan}
\end{split}
\end{equation}
and
\begin{equation}
\begin{split}
&\eta_{\rm XPM,cc}(f_i) = \frac{32}{27}\frac{\gamma^2}{B_k^2}\left(\frac{P_k}{P_i}\right)^2 {T'_i}^2 \sum_{\substack{l, l' \in \{0, 1\}}} \left( \frac{-\tilde{T}'_k}{T'_k} \right)^{l+l'} \\
&\times \frac{B_k\kappa_{l,k} \kappa_{l',k}}{\phi_{i,k}\sqrt{\tilde{\alpha}_{l,k} \tilde{\alpha}_{l',k}}} \sum_{n=1}^{N_{\rm s} - 1} 2\left(N_{\rm s} - n\right) \left[\sign(\phi_{i,k}) \pi \vphantom{\frac{2\sqrt{\tilde{\alpha}_{l,k} \tilde{\alpha}_{l',k}}\sin\left(j n\phi_{i,k} L\frac{B_i}{2}\right)}{nL\left(\tilde{\alpha}_{l,k}\tilde{\alpha}_{l',k}+\phi_{i,k}^2\frac{B_i^2}{4}\right)}} \right. \\
&\left. \times e^{-nL\sqrt{\tilde{\alpha}_{l,k} \tilde{\alpha}_{l',k}}} + \frac{2\sqrt{\tilde{\alpha}_{l,k} \tilde{\alpha}_{l',k}}\sin\left(j n\phi_{i,k} L\frac{B_i}{2}\right)}{nL\left(\tilde{\alpha}_{l,k}\tilde{\alpha}_{l',k}+\phi_{i,k}^2\frac{B_i^2}{4}\right)} \right],
\label{eq:XPM_GN_closed_multispan}
 \end{split}
\end{equation}
where $\tilde{\alpha}_{l,i}$, $\tilde{\alpha}_{l',i}$, $\kappa_{l,i}$, and $\kappa_{l',i}$ can be found in~\cite[Eqs.~(15)~and~(16)]{Buglia_JLT_lowloss}. $\phi_{i}$ and $\phi_{i,k}$ can be found in~\cite[Eqs.~(26)~and~(27)]{Min_JLT} with support of $\beta_4$. The proof of Eqs~\eqref{eq:SPM_GN_closed_multispan}~and~\eqref{eq:XPM_GN_closed_multispan} are given respectively in Appendix~\ref{appC:SPM_multispan}~and~\ref{appD:XPM_multispan}.

\section{Results}
\label{sec:Results}
This section describes the numerical validation of the final CFM where closed-form expression of FWM contributions shown in Eq.~\eqref{eq:FWM_GN_closed} in conjunction with closed-form expressions of SPM/XPM coherent contributions shown in Eqs.~\eqref{eq:SPM_GN_closed_multispan}~and~\eqref{eq:XPM_GN_closed_multispan} are included. The nonlinear interference coefficient ($\eta_{\rm NLI}$) and nonlinear interference SNR ($\SNR_{\rm NLI}$) evaluated using the proposed CFM was compared with SSFM simulations, ISRS GN model in integral form, and CFM with incoherent assumptions on SPM, XPM, and FWM.

\subsection{Transmission Setup}
\label{Transmission Setup}

The baseline transmission system, over which the derived expressions are validated, consists of a 1$\times$80~km to 10$\times$80~km WDM transmission link with $N_{\rm ch} = 161$ channels spaced by 100~GHz and centred at the zero-dispersion wavelength of 1302.3~nm. Each channel was modulated at the symbol rate of 96~GBd. This resulted in a total bandwidth of 16.1~THz (90.6~nm), ranging from 1258.6~nm to 1349.2~nm, corresponding to transmission over the O-band. The channels were transmitted using a single-mode
fibre (SMF) where the number of spans is varied as described in the next sections. It is assumed that each amplifier fully compensates for the span losses (the transparent link assumption). A spectrally uniform input launch power profile was used. Realistic wavelength-dependent attenuation and dispersion profile, shown in Fig.~\ref{fig:att_dispersion}, and Raman gain spectrum were used~\cite{Min_JLT}. The nonlinearity coefficient was $\gamma = 2$~W\textsuperscript{-1}km\textsuperscript{-1}. These parameters are summarised in Table~\ref{table:1}.

\begin{figure}

\begin{tikzpicture}
    \begin{axis}
    [
    width=0.45\textwidth-1.85626pt,height=6cm,
    grid=both,
    legend style={fill opacity=1, draw opacity=1, text opacity=1, at={(0.5,0.3)}, anchor=south west, draw=black, font=\small},
    xlabel={Wavelength~[nm]},
    ylabel={Dispersion~[ps/nm/km]},
    ylabel near ticks,
    xlabel near ticks,
    ylabel shift = -2 pt,
    xlabel shift = -2 pt,
    ytick distance=1,
    ymin=-5,
    ymax=5,
    ytick={-4,-2,0,2,4},
    yticklabels={-4,-2,0,2,4},
    xmin=1255,
    xmax=1355,
    clip marker paths=true,
    tick label style={
        /pgf/number format/.cd,
        fixed,
        precision=3,
        use comma=false,
        1000 sep={},
        set decimal separator={.}
    },
    scaled ticks=false,
    axis y line*=left,
    axis x line*=box,
    ]

        \addplot[green1,thick,solid,mark=none,opacity=0.5] table[x=wvec,y=d] {Data/att_dispersion.tsv};
        \label{d}

        \addplot[green1,thick,dashed,mark=none] table[x=wvec,y=fitd] {Data/att_dispersion.tsv};
        \label{fitd}

        \addplot[green1,thick,dotted,mark=none] table[x=wvec,y=fitd2] {Data/att_dispersion.tsv};
        \label{fitd2}

    \end{axis}

    \begin{axis}
    [
    legend style={
        at={(0, 1.02)},
        anchor=south west,
    },
    legend columns=4,
    width=0.45\textwidth-1.85626pt,height=6cm,
    axis y line*=right,
    axis x line=none,
    ymin=0.265,
    ymax=0.415,
    ytick={0.28,0.31,0.34,0.37,0.4},
    yticklabels={0.28,0.31,0.34,0.37,0.40},
    xmin=1255,
    xmax=1355,
    ylabel={Attenuation~[dB/km]},
    ylabel near ticks,
    ylabel shift = -2 pt,
    tick label style={
        /pgf/number format/.cd,
        fixed,
        precision=3,
        use comma=false,
        1000 sep={},
        set decimal separator={.}
    },
    scaled ticks=false,
    ]
        \addlegendimage{/pgfplots/refstyle=d}\addlegendentry{$D(\lambda)$}
        \addlegendimage{/pgfplots/refstyle=fitd}\addlegendentry{$D_{\rm 1st}(\lambda)$}
        \addlegendimage{/pgfplots/refstyle=fitd2}\addlegendentry{$D_{\rm 2nd}(\lambda)$}
    
        \addplot[orange1,thick,solid,mark=none] table[x=wvec,y=alpha] {Data/att_dispersion.tsv};
        \addlegendentry{$\alpha(\lambda)$}
    
    \end{axis}

\end{tikzpicture}

\caption{Simulated fibre attenuation $\alpha(\lambda)$, dispersion profile $D(\lambda)$ and its fitting curves: $D_{\rm 1st}(\lambda)$ considering $\beta_2$ and $\beta_3$ terms, and $D_{\rm 2nd}(\lambda)$ which also includes $\beta_4$ term.}
\label{fig:att_dispersion}

\end{figure}
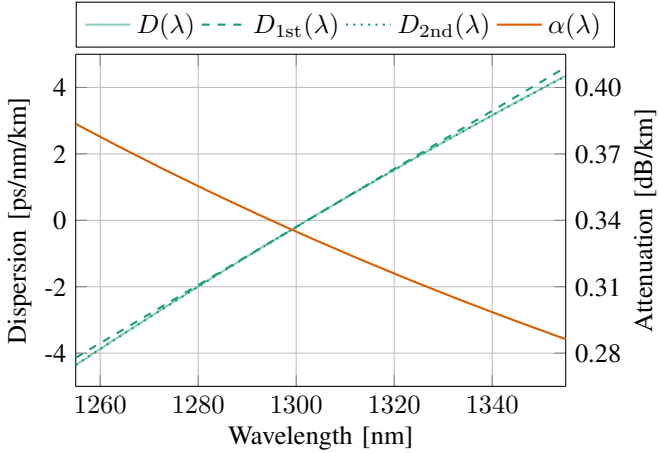 
\begin{table}[!t]
    \centering
    \caption{System Parameters}
    \begin{tabular}{ccc}
        \hline
        \textbf{Parameters} & \textbf{Unit} & \textbf{Value} \\
        \hline
        Reference wavelength $(\lambda_c)$ & nm & 1302.3 \\
        Dispersion $(D)$ & ps/nm/km & 0 \\
        Dispersion slope $(S)$ & ps/nm\textsuperscript{2}/km & 0.087 \\
        Dispersion curvature $(\dot{S})$ & ps/nm\textsuperscript{3}/km & -9.714$\cdot$10\textsuperscript{-5} \\
        NL coefficient $(\gamma)$ & 1/W/km & 2 \\
        Symbol rate & GBd & 96 \\
        Channel spacing & GHz & 100 \\
        Number of channels & - & 161 \\
        Modulation & - & Gaussian \\
        \hline
    \end{tabular}
    \label{table:1}
\end{table}

\subsection{Numerical Validation}
\label{Numerical Validation}

\begin{figure*}

\begin{tikzpicture}

    \begin{groupplot}
    [
    legend columns=3,
    width=0.4\textwidth-6.47649/3pt,height=6cm,
    grid=both,
    legend style={fill opacity=1, draw opacity=1, text opacity=1, at={(0.02,0.02)}, anchor=south west, draw=black, font=\small},
    ylabel near ticks,
    xlabel near ticks,
    ylabel shift = -2 pt,
    xlabel shift = -2 pt,
    clip marker paths=true,
    group style={group size=3 by 1, horizontal sep=0.1cm,xlabels at=edge bottom,ylabels at=edge left},
    ]
    
    \nextgroupplot[
    ylabel={$\eta_{\rm NLI}$~[dB~($1\cdot \text{W}^{-2}$)]},
    ymin=32,
    ymax=46,
    ytick={32,34,36,38,40,42,44,46},
    yticklabels={32,34,36,38,40,42,44,46},
    xmin=1255,
    xmax=1355,
    xlabel={Wavelength [nm]},
    tick label style={
        /pgf/number format/.cd,
        fixed,
        precision=3,
        use comma=false,
        1000 sep={},
        set decimal separator={.}
    },
    scaled ticks=false,
    ]

        \addplot[matlab1,thick,solid,no marks,forget plot] table[x=wvec,y=eta_ssfm_-2] {Data/eta_1span.tsv};
        \addplot[matlab1,thick,dashed,no marks,forget plot] table[x=wvec,y=eta_integral_-2] {Data/eta_1span.tsv};
        \addplot[matlab1,thick,dotted,no marks,forget plot] table[x=wvec,y=eta_cf_-2] {Data/eta_1span.tsv};
        
        \addlegendimage{black,thick,solid}
        \addlegendentry{SSFM}
        \addlegendimage{black,thick,dashed}
        \addlegendentry{Integral}
        \addlegendimage{black,thick,dotted}
        \addlegendentry{CFM}

        \node[anchor=south west] at (rel axis cs:0.2,0.2) {(a) -2~dBm};

    \nextgroupplot[
    ymin=32,
    ymax=46,
    ytick={32,34,36,38,40,42,44,46},
    yticklabels={},
    xmin=1255,
    xmax=1355,
    xlabel={Wavelength~[nm]},
    tick label style={
        /pgf/number format/.cd,
        fixed,
        precision=3,
        use comma=false,
        1000 sep={},
        set decimal separator={.}
    },
    scaled ticks=false,
    ]

        \addplot[matlab2,thick,solid,no marks,forget plot] table[x=wvec,y=eta_ssfm_0] {Data/eta_1span.tsv};
        \addplot[matlab2,thick,dashed,no marks,forget plot] table[x=wvec,y=eta_integral_0] {Data/eta_1span.tsv};
        \addplot[matlab2,thick,dotted,no marks,forget plot] table[x=wvec,y=eta_cf_0] {Data/eta_1span.tsv};

        \addlegendimage{black,thick,solid}
        \addlegendentry{SSFM}
        \addlegendimage{black,thick,dashed}
        \addlegendentry{Integral}
        \addlegendimage{black,thick,dotted}
        \addlegendentry{CFM}

        \node[anchor=south west] at (rel axis cs:0.2,0.2) {(b) 0~dBm};

    \nextgroupplot[
    ymin=32,
    ymax=46,
    ytick={32,34,36,38,40,42,44,46},
    yticklabels={},
    xmin=1255,
    xmax=1355,
    xlabel={Wavelength~[nm]},
    tick label style={
        /pgf/number format/.cd,
        fixed,
        precision=3,
        use comma=false,
        1000 sep={},
        set decimal separator={.}
    },
    scaled ticks=false,
    ]

        \addplot[matlab3,thick,solid,no marks,forget plot] table[x=wvec,y=eta_ssfm_2] {Data/eta_1span.tsv};
        \addplot[matlab3,thick,dashed,no marks,forget plot] table[x=wvec,y=eta_integral_2] {Data/eta_1span.tsv};
        \addplot[matlab3,thick,dotted,no marks,forget plot] table[x=wvec,y=eta_cf_2] {Data/eta_1span.tsv};

        \addlegendimage{black,thick,solid}
        \addlegendentry{SSFM}
        \addlegendimage{black,thick,dashed}
        \addlegendentry{Integral}
        \addlegendimage{black,thick,dotted}
        \addlegendentry{CFM}

        \node[anchor=south west] at (rel axis cs:0.2,0.2) {(c) 2~dBm};

    \end{groupplot}

\end{tikzpicture}

\caption{Nonlinear interference coefficient ($\eta_{\rm NLI}$) for 161-channel single-span transmission centred at zero-dispersion wavelength with per-channel launch power of (a) -2~dBm, (b) 0~dBm, and (c) 2~dBm. The results from SSFM, ISRS GN integral model, and proposed CFM models are compared. The optimal flat launch power per channel is found at -0.52~dBm with an ideal lumped amplification with 5~dB noise figure.}
\label{fig:eta_1span}

\end{figure*}
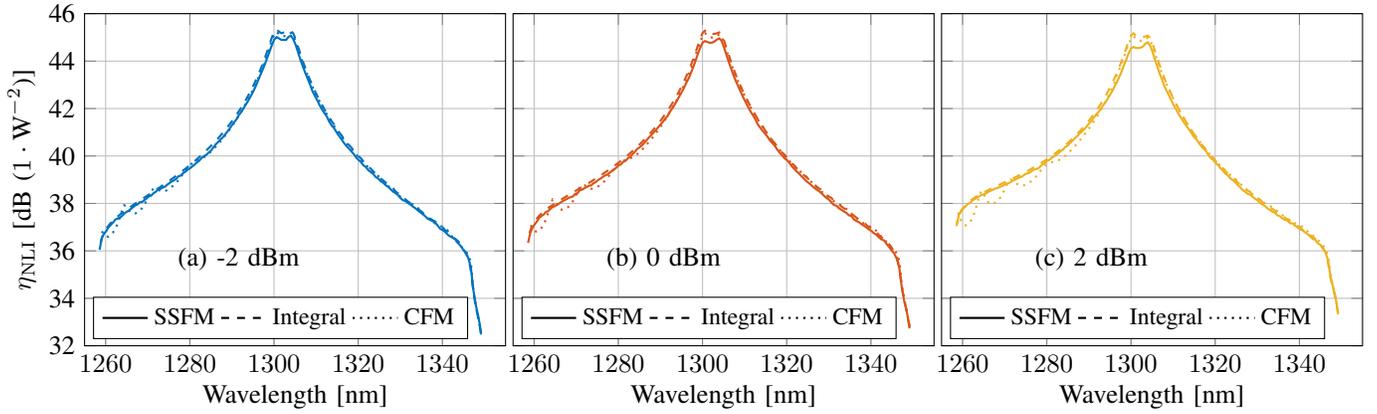 %
\begin{figure}

\begin{tikzpicture}
    \begin{groupplot}
    [
    axis on top,
    legend columns=2,
    width=0.31\textwidth-9.8909/2pt,height=0.31\textwidth-9.8909/2pt,
    legend style={fill opacity=1, draw opacity=1, text opacity=1, at={(0.47,0.98)}, anchor=north west, draw=black, font=\small},
    xlabel={$f_1$~[THz]},
    ylabel near ticks,
    xlabel near ticks,
    ylabel shift = -2 pt,
    xlabel shift = -2 pt,
    ytick distance=1,
    clip marker paths=true,
    tick label style={
        /pgf/number format/.cd,
        fixed,
        precision=3,
        use comma=false,
        1000 sep={},
        set decimal separator={.},
    },
    scaled ticks=false,
    axis x line*=box,
    group style={group size=2 by 1, horizontal sep=0.1cm,xlabels at=edge bottom,ylabels at=edge left},
    ]

    \nextgroupplot[
    ylabel={$f_2$~[THz]},
    ymin=-8.05,
    ymax=8.05,
    ytick={-6,-2,2,6},
    yticklabels={-6,-2,2,6},
    xmin=-8.05,
    xmax=8.05,
    xtick={-6,-2,2,6},
    xticklabels={-6,-2,2,6},
    colorbar horizontal,
    colormap/viridis,
    point meta min=-125,
    point meta max=-35,
    colorbar style={
        xlabel={$P_{\rm FWM}$~[dBm]},
        xlabel near ticks,
        xlabel shift = -2 pt,
        at={(0,1.02)},
        anchor=south west,
        width=
        2*\pgfkeysvalueof{/pgfplots/parent axis width}+
        \pgfkeysvalueof{/pgfplots/group/horizontal sep},
        xticklabel pos=upper,
    },
    ]
        \addplot graphics [ymin=-8.05,ymax=8.05,xmin=-8.05,xmax=8.05] {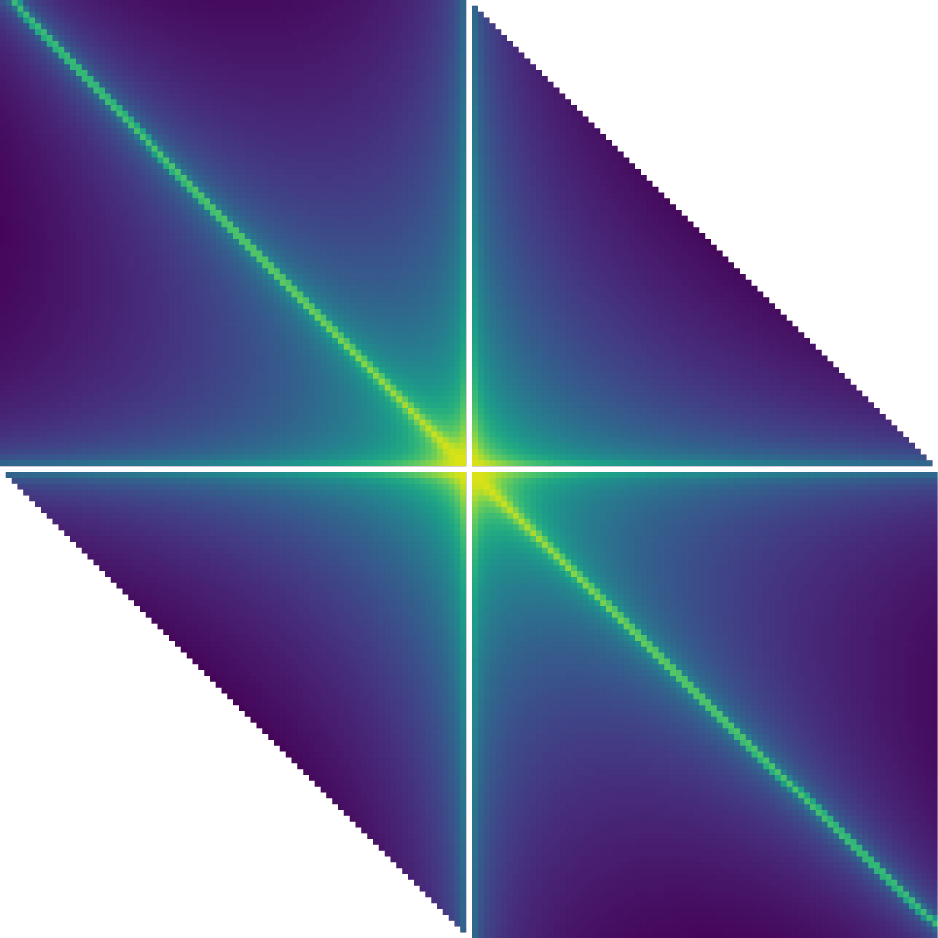};

    \nextgroupplot[
    ymin=-8.05,
    ymax=8.05,
    ytick={-6,-2,2,6},
    yticklabels={},
    xmin=-8.05,
    xmax=8.05,
    xtick={-6,-2,2,6},
    xticklabels={-6,-2,2,6},
    ]
        \addplot graphics [ymin=-8.05,ymax=8.05,xmin=-8.05,xmax=8.05] {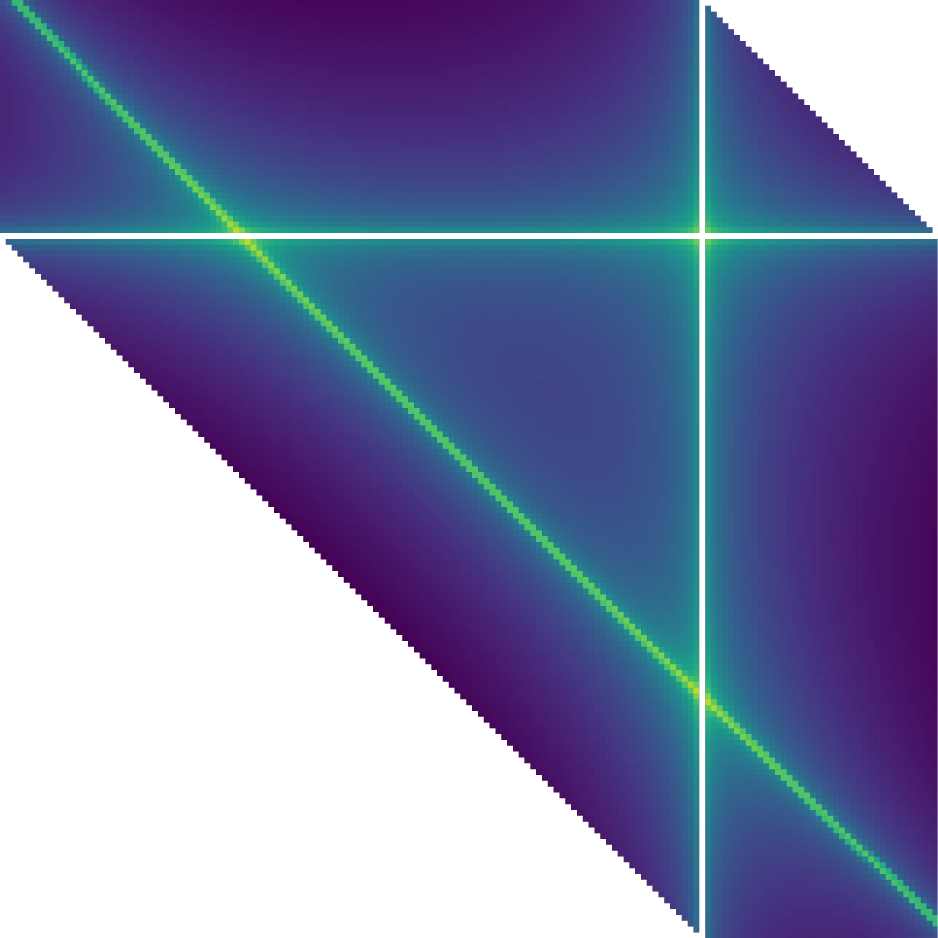};

    \end{groupplot}
    
\end{tikzpicture}

\caption{FWM contributions evaluated by the proposed CFM at channel $i=81$ (centre channel) and channel $i=121$, in 161-channel transmission with -2~dBm launch power per channel.}
\label{fig:fwm_err}

\end{figure} 
CFM uses~\cite{Buglia_JLT_lowloss} formula for SPM and XPM but were expanded to support the fourth-order GVD parameter $\beta_4$, and transmission over multiple spans in O band by introducing new coherent factors for SPM and XPM given by Eqs.~\eqref{eq:SPM_GN_closed_multispan}~and~\eqref{eq:XPM_GN_closed_multispan}. Additionally, the CFM uses a FWM contribution calculated from Eq.~\eqref{eq:FWM_GN_closed}. To verify the accuracy of the proposed closed-form expression, it was compared with the ISRS GN model in integral form and with SSFM simulations~\cite{Min_JLT} using Gaussian constellations. The SSFM simulation considers $2^{16}$ random Gaussian symbols per polarisation per channel with two samples per symbol and a root-raised-cosine filter with a roll-off factor of 0.01\%. Step sizes for the SSFM simulation were optimised using the local-error method~\cite{sinkin2003optimization} using a small goal error value of $\delta_G = 10^{-9}$ to ensure accurate results. The integral model was solved using an average of 2 steps per km ($\bar{N}_{M}$) and 500 Riemann samples ($N_{R}$).

A demonstration of the FWM evaluation using the proposed CFM is shown in Fig.~\ref{fig:fwm_err}. The FWM contributions, $P_{\rm FWM}\left(f_i\right) = \eta_{\rm FWM}\left(f_i\right)P_i^3$, of all necessary frequency triplets $\left(f_j, f_k, f_m\right)$ in $\Omega$ are calculated. The two axes marked in white correspond to SPM and XPM contributions. The remaining colourful regions correspond to FWM triplets. It is shown that FWM contributions are more significant near axes $f_1 = f_i$, $f_2 = f_i$, and $f_1+f_2 = 0$. The same results were obtained using integral ISRS GN model, showing a $P_{\rm FWM}$ mean squared error between CFM and integral model of 0.34~dB and 0.35~dB at channel $i=81$ and $i=121$, respectively.

The per-channel nonlinear coefficient for the 161-channel single-span case is plotted in Fig.~\ref{fig:eta_1span}. To verify the accuracy of the CFM, the channel-wise absolute error relative to the SSFM simulation and integral model are defined as $\left|\eta_{\rm NLI,CFM}^{\rm dB}-\eta_{\rm NLI,SSFM}^{\rm dB}\right|$ and $\left|\eta_{\rm NLI,GN}^{\rm dB}-\eta_{\rm NLI,SSFM}^{\rm dB}\right|$, respectively. The maximum nonlinear coefficient error across the entire signal bandwidth between CFM and SSFM simulation is 0.86~dB at 1261.2~nm being for the case with 2~dBm launch power per channel shown in Fig.~\ref{fig:eta_1span}(c). This corresponds to a total launch power of 24~dBm, well above the optimum. The maximum nonlinear coefficient error is followed by 0.68~dB at 1258.6~nm for the scenario with 0~dBm per channel in Fig.~\ref{fig:eta_1span}(b) and 0.63~dB error at the same wavelength with -2~dBm per channel in Fig.~\ref{fig:eta_1span}(a). Within the near–zero-dispersion region (i.e., $|D|\leq 1$~ps/nm/km, approximately 1290 and 1314~nm), the maximum error is 0.45~dB at 1300.6~nm, which occurs for the channel with the highest nonlinear coefficient. The maximum nonlinear coefficient error within near-zero-dispersion region is 0.33~dB for the scenario with 0~dBm per channel and 0.21~dB error with -2~dBm per channel both at 1300.6~nm. For all the cases, good agreement is found between CFM and SSFM models, with just 0.12, 0.16, and 0.22~dB mean absolute errors being observed across the entire bandwidth. However, in channels that are located below 1280~nm, ripples are observed in the nonlinear coefficient, which is where the maximum nonlinear coefficient error occurs. This ripples are caused by the inclusion of $\beta_4$ in the phase-mismatch term. Results for optical powers below -2~dBm are very similar to -2~dBm case and, therefore, are not included.

For the same 161-channel scenario, we also include, for comparison, the results obtained using the integral model in~\cite{Min_JLT}. Estimation of the nonlinear coefficient using the integral model is shown in Fig.~\ref{fig:eta_1span}. The maximum channel errors of $\eta_{\rm NLI}$ in dB are 0.32, 0.44, and 0.57~dB for -2, 0, and 2~dBm scenarios, respectively, and all at the same wavelength of 1300.6~nm. The average channel errors are 0.14, 0.16, and 0.15~dB for the same scenarios. The error between the GN model and the SSFM simulation is expected to be larger for high launch powers because of the first-order regular perturbation assumption underlying the derivation of the GN model, which results in an NLI overestimation. Although the CFM shows a smaller error than the integral model, this does not imply superior accuracy. Both methods overestimate NLI due to the first-order regular perturbation assumption. The additional assumptions built into the CFM (see Section~\ref{sec:closedform_expression}) further reduce its NLI estimate, so the resulting error appears to be smaller.

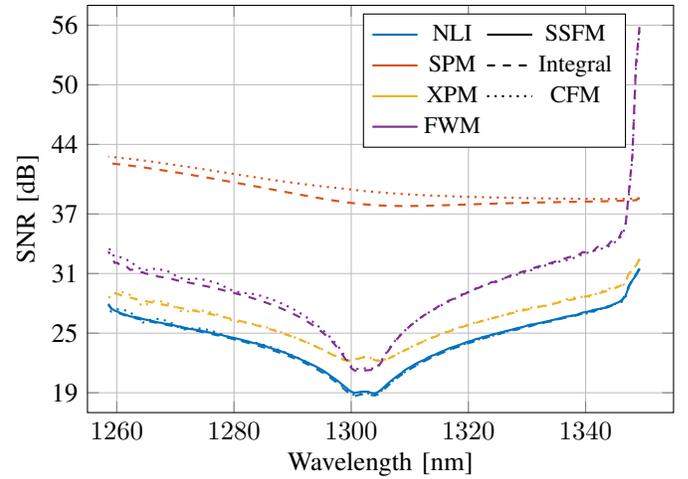
\begin{figure}

\begin{tikzpicture}
    \begin{axis}
    [
    legend columns=2,
    width=0.52\textwidth-1.63326pt,height=7cm,
    grid=both,
    legend style={fill opacity=1, draw opacity=1, text opacity=1, at={(0.47,0.98)}, anchor=north west, draw=black, font=\small},
    xlabel={Wavelength~[nm]},
    ylabel={SNR~[dB]},
    ylabel near ticks,
    xlabel near ticks,
    ylabel shift = -2 pt,
    xlabel shift = -2 pt,
    ytick distance=1,
    ymin=17,
    ymax=58,
    ytick={19,25,31,37,44,50,56},
    yticklabels={19,25,31,37,44,50,56},
    xmin=1255,
    xmax=1355,
    clip marker paths=true,
    tick label style={
        /pgf/number format/.cd,
        fixed,
        precision=3,
        use comma=false,
        1000 sep={},
        set decimal separator={.}
    },
    scaled ticks=false,
    axis x line*=box,
    ]

        \addplot[matlab1,thick,solid,mark=none,forget plot] table[x=wvec,y=SNR_SSFM] {Data/snr_1span_-2.tsv};  
        \addplot[matlab1,thick,dashed,mark=none,forget plot] table[x=wvec,y=SNR_NLI_integral] {Data/snr_1span_-2.tsv};
        \addplot[matlab1,thick,dotted,mark=none,forget plot] table[x=wvec,y=SNR_NLI] {Data/snr_1span_-2.tsv};

        \addplot[matlab2,thick,dashed,mark=none,forget plot] table[x=wvec,y=SNR_SPM_integral] {Data/snr_1span_-2.tsv};
        \addplot[matlab2,thick,dotted,mark=none,forget plot] table[x=wvec,y=SNR_SPM] {Data/snr_1span_-2.tsv};

        \addplot[matlab3,thick,dashed,mark=none,forget plot] table[x=wvec,y=SNR_XPM_integral] {Data/snr_1span_-2.tsv};
        \addplot[matlab3,thick,dotted,mark=none,forget plot] table[x=wvec,y=SNR_XPM] {Data/snr_1span_-2.tsv};

        \addplot[matlab4,thick,dashed,mark=none,forget plot] table[x=wvec,y=SNR_FWM_integral] {Data/snr_1span_-2.tsv};
        \addplot[matlab4,thick,dotted,mark=none,forget plot] table[x=wvec,y=SNR_FWM] {Data/snr_1span_-2.tsv};

        \addlegendimage{matlab1,thick,solid}
        \addlegendentry{NLI}
        \addlegendimage{black,thick,solid}
        \addlegendentry{SSFM}
        \addlegendimage{matlab2,thick,solid}
        \addlegendentry{SPM}
        \addlegendimage{black,thick,dashed}
        \addlegendentry{Integral}
        \addlegendimage{matlab3,thick,solid}
        \addlegendentry{XPM}
        \addlegendimage{black,thick,dotted}
        \addlegendentry{CFM}
        \addlegendimage{matlab4,thick,solid}
        \addlegendentry{FWM}
         
    \end{axis}

\end{tikzpicture}

\caption{SPM, XPM and FWM contributions in terms of SNR as a function of wavelength for launch power of -2~dBm per channel in a single-span transmission.}
\label{fig:snr_1span_-2}

\end{figure}

A detailed presentation of nonlinear SNR, i.e., $\SNR_\NLI$, due to different contributions is shown in Fig.~\ref{fig:snr_1span_-2}, evaluated by the metric $\left|\SNR_{\rm NLI,GN}^{\rm dB}-\SNR_{\rm NLI,SSFM}^{\rm dB}\right|$ and $\left|\SNR_{\rm NLI,CFM}^{\rm dB}-\SNR_{\rm NLI,SSFM}^{\rm dB}\right|$. It shows that FWM is dominant between 1299.4 and 1305.1~nm. The mean absolute difference of FWM contribution in $\SNR_{\rm NLI}$ between the CFM and the integral model is 0.27~dB and the maximum difference is 0.92~dB at 1260.7~nm. The reason for the higher error within the negative dispersion spectral region is the same as that mentioned above, i.e., it is caused by the inclusion of $\beta_4$ into the phase-mismatch. The discrepancies in the calculated SPM and XPM contributions are 0.79 and 0.10~dB, respectively, where the relatively large difference in SPM is caused by the approximated circular integration domain in~\cite{Buglia_JLT_lowloss}. It leads to an overall average of 0.13~dB SNR difference in the NLI calculation, which still provides an accurate and fast evaluation of the NLI noise.

To verify the accuracy of the new coherent factor $\epsilon$ Eqs.~\eqref{eq:SPM_GN_closed_multispan}~and~\eqref{eq:XPM_GN_closed_multispan} are used for the case where $N_{\rm s}=10$, with other parameters being kept unchanged. The $\SNR_{\rm NLI}$ was compared with the integral model and the incoherent CFM - where all NLI contributions are assumed to accumulate incoherently - using $\left|\SNR_{\rm NLI,CFM}^{\rm dB}-\SNR_{\rm NLI,GN}^{\rm dB}\right|$. The resulting $\SNR_{\rm NLI}$ variation with distance is shown in Fig.~\ref{fig:snr_10span_spm_xpm}. For the SPM contribution, which is the term affected the most by coherence, the CFM result differs by a mean value of 0.13~dB from that of the integral model. The incoherent CFM gives an 11.3~dB error at the zero-dispersion wavelength, and an average 3.52~dB across the entire bandwidth. For the XPM contribution, as shown in Fig.~\ref{fig:snr_10span_spm_xpm}, the coherence only affects channels near the zero-dispersion region. Eq.~\eqref{eq:XPM_GN_closed_multispan} reduces the 3.2~dB gap from the incoherent CFM to 0.04~dB. It still has an average 0.38~dB difference, mainly caused by the ripple at shorter wavelengths. Similarly to Fig.~\ref{fig:snr_1span_-2}, each nonlinear contribution in terms of $\SNR_{\rm NLI}$ is plotted in the same 10-span simulation, as shown in Fig.~\ref{fig:snr_10span_-2}, where it is observed that XPM is dominant over the entire band.

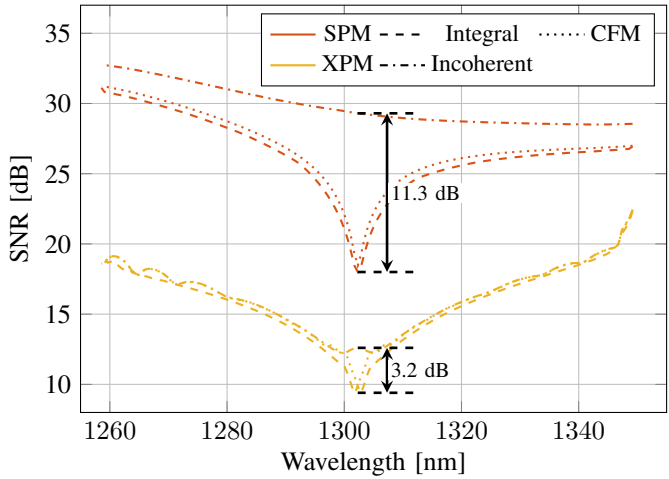
\begin{figure}

\begin{tikzpicture}
    \begin{axis}
    [
    legend columns=3,
    width=0.52\textwidth-1.63326pt,height=7cm,
    grid=both,
    legend style={fill opacity=1, draw opacity=1, text opacity=1, at={(0.98,0.98)}, anchor=north east, draw=black, font=\small},
    xlabel={Wavelength~[nm]},
    ylabel={SNR~[dB]},
    ylabel near ticks,
    xlabel near ticks,
    ylabel shift = -2 pt,
    xlabel shift = -2 pt,
    ytick distance=1,
    ymin=8,
    ymax=37,
    ytick={10,15,20,25,30,35},
    yticklabels={10,15,20,25,30,35},
    xmin=1255,
    xmax=1355,
    clip marker paths=true,
    tick label style={
        /pgf/number format/.cd,
        fixed,
        precision=3,
        use comma=false,
        1000 sep={},
        set decimal separator={.}
    },
    scaled ticks=false,
    axis x line*=box,
    ]

        \addplot[matlab2,thick,dashed,mark=none,forget plot] table[x=wvec,y=SNR_SPM_integral] {Data/snr_10span_spm_xpm.tsv};
        \addplot[matlab2,thick,dotted,mark=none,forget plot] table[x=wvec,y=SNR_SPM] {Data/snr_10span_spm_xpm.tsv};
        \addplot[matlab2,thick,dashdotted,mark=none,forget plot] table[x=wvec,y=SNR_SPM2] {Data/snr_10span_spm_xpm.tsv};

        \addplot[matlab3,thick,dashed,mark=none,forget plot] table[x=wvec,y=SNR_XPM_integral] {Data/snr_10span_spm_xpm.tsv};
        \addplot[matlab3,thick,dotted,mark=none,forget plot] table[x=wvec,y=SNR_XPM] {Data/snr_10span_spm_xpm.tsv};
        \addplot[matlab3,thick,dashdotted,mark=none,forget plot] table[x=wvec,y=SNR_XPM2] {Data/snr_10span_spm_xpm.tsv};

        \addlegendimage{matlab2,thick,solid}
        \addlegendentry{SPM}
        \addlegendimage{black,thick,dashed}
        \addlegendentry{Integral}
        \addlegendimage{black,thick,dotted}
        \addlegendentry{CFM}
        \addlegendimage{matlab3,thick,solid}
        \addlegendentry{XPM}
        \addlegendimage{black,thick,dashdotted}
        \addlegendentry{Incoherent}

        \draw [dashed,line width = 1] (axis cs:1302.3,18)--(axis cs:1302.3+10,18);
        \draw [dashed,line width = 1] (axis cs:1302.3,29.3)--(axis cs:1302.3+10,29.3);
        \draw [stealth-stealth,line width = 1] (axis cs:1302.3+5,18)--(axis cs:1302.3+5,29.3) node[midway,fill=white,inner sep=1pt,font=\footnotesize,anchor=west] {11.3~dB};

        \draw [dashed,line width = 1] (axis cs:1302.3,9.4)--(axis cs:1302.3+10,9.4);
        \draw [dashed,line width = 1] (axis cs:1302.3,12.6)--(axis cs:1302.3+10,12.6);
        \draw [stealth-stealth,line width = 1] (axis cs:1302.3+5,9.4)--(axis cs:1302.3+5,12.6) node[midway,fill=white,inner sep=1pt,font=\footnotesize,anchor=west] {3.2~dB};
         
    \end{axis}

\end{tikzpicture}

\caption{SPM and XPM contributions in terms of SNR as a function of wavelength for launch power of -2~dBm per channel in a 10-span transmission. Incoherent CFM results were included for comparison.}
\label{fig:snr_10span_spm_xpm}

\end{figure}
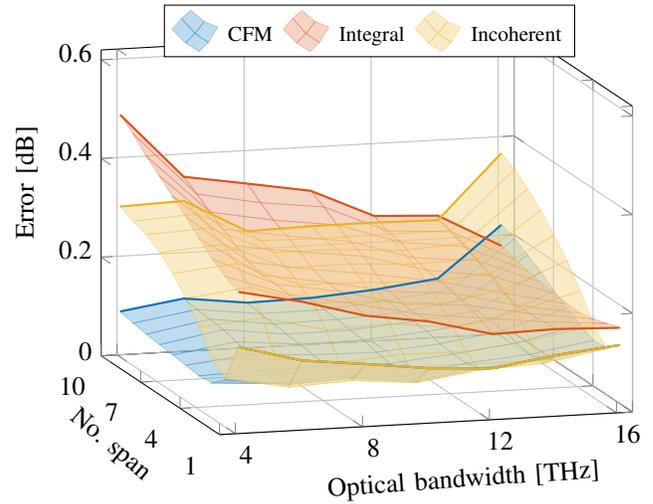
\begin{figure}
\centering
\begin{tikzpicture}
\begin{axis}[
    unbounded coords=jump,view={74}{15}, grid=both,
    legend cell align=left,
    legend style={font=\footnotesize, column sep=2pt,at={(rel axis cs:0,.65,1.1)}, anchor=north},
    legend columns=3,
    width=0.5\textwidth-12.04718pt,height=7cm,
    x label style={rotate=-40,yshift=.5cm},
    y label style={rotate=4,yshift=.2cm},
    xlabel={No. span},
    ylabel={Optical bandwidth [THz]},
    zlabel={Error [dB]},
    x tick label style = {text width = 1.0cm, align = center, rotate = 0},
    y tick label style = {text width = 1.0cm, align = center, rotate = 0},
    x dir=reverse,
    zmin=0,zmax=0.62,
    ymin=3.5,ymax=16.5,
    xmin=1,xmax=10,
    ytick={4,8,12,16},
    yticklabels={4,8,12,16},
    xtick distance=5,
    xtick={1,4,7,10},
    xticklabels={1,4,7,10},
    ztick distance=5,
    ztick={0,0.2,0.4,0.6},
    zticklabels={0,0.2,0.4,0.6},
]

\addplot3[surf,matlab1!50,opacity=0.5,mesh/rows=7,mesh/cols=10,faceted color=matlab1!80,line width=0.3pt] table [x=Nspan_all,y=Nch_all,z=err] {Data/err.tsv};
\addplot3[surf,matlab2!50,opacity=0.5,mesh/rows=7,mesh/cols=10,faceted color=matlab2!80,line width=0.3pt] table [x=Nspan_all,y=Nch_all,z=err_integral] {Data/err.tsv};
\addplot3[surf,matlab3!50,opacity=0.5,mesh/rows=7,mesh/cols=10,faceted color=matlab3!80,line width=0.3pt] table [x=Nspan_all,y=Nch_all,z=err_inc] {Data/err.tsv};

\addplot3[matlab1,thick,forget plot] table [x=Nspan_1,y=Nch_1,z=err_1] {Data/err_end.tsv};
\addplot3[matlab1,thick,forget plot] table [x=Nspan_10,y=Nch_1,z=err_10] {Data/err_end.tsv};

\addplot3[matlab2,thick,forget plot] table [x=Nspan_1,y=Nch_1,z=err_integral_1] {Data/err_end.tsv};
\addplot3[matlab2,thick,forget plot] table [x=Nspan_10,y=Nch_1,z=err_integral_10] {Data/err_end.tsv};

\addplot3[matlab3,thick,forget plot] table [x=Nspan_1,y=Nch_1,z=err_inc_1] {Data/err_end.tsv};
\addplot3[matlab3,thick,forget plot] table [x=Nspan_10,y=Nch_1,z=err_inc_10] {Data/err_end.tsv};

\addlegendentry{CFM};
\addlegendimage{surf,matlab1!50,opacity=0.5,faceted color=matlab1!80,line width=0.3pt};
\addlegendentry{Integral};
\addlegendimage{surf,matlab2!50,opacity=0.5,faceted color=matlab2!80,line width=0.3pt};
\addlegendentry{Incoherent};
\addlegendimage{surf,matlab3!50,opacity=0.5,faceted color=matlab3!80,line width=0.3pt};

\end{axis}
\end{tikzpicture}
\caption{Mean absolute error of the proposed CFM, integral, and incoherent CFM compared to SSFM as a function of different number of spans and optical bandwidth.}
\label{fig:err}
\end{figure}

\begin{figure}
\centering
\begin{tikzpicture}
\begin{axis}[
    unbounded coords=jump,view={74}{15}, grid=both,
    legend cell align=left,
    legend style={font=\footnotesize, column sep=2pt,at={(rel axis cs:0,.65,1.1)}, anchor=north},
    legend columns=3,
    width=0.5\textwidth-12.04718pt,height=7cm,
    x label style={rotate=-40,yshift=.5cm},
    y label style={rotate=4,yshift=.2cm},
    xlabel={No. span},
    ylabel={Optical bandwidth [THz]},
    zlabel={Error [dB]},
    x tick label style = {text width = 1.0cm, align = center, rotate = 0},
    y tick label style = {text width = 1.0cm, align = center, rotate = 0},
    x dir=reverse,
    zmin=0,zmax=1.7,
    ymin=3.5,ymax=16.5,
    xmin=1,xmax=10,
    ytick={4,8,12,16},
    yticklabels={4,8,12,16},
    xtick distance=5,
    xtick={1,4,7,10},
    xticklabels={1,4,7,10},
    ztick distance=5,
    ztick={0,0.5,1,1.5},
    zticklabels={0,0.5,1,1.5},
]

\addplot3[surf,matlab1!50,opacity=0.5,mesh/rows=7,mesh/cols=10,faceted color=matlab1!80,line width=0.3pt] table [x=Nspan_all,y=Nch_all,z=err_max] {Data/err_max.tsv};
\addplot3[surf,matlab2!50,opacity=0.5,mesh/rows=7,mesh/cols=10,faceted color=matlab2!80,line width=0.3pt] table [x=Nspan_all,y=Nch_all,z=err_integral_max] {Data/err_max.tsv};
\addplot3[surf,matlab3!50,opacity=0.5,mesh/rows=7,mesh/cols=10,faceted color=matlab3!80,line width=0.3pt] table [x=Nspan_all,y=Nch_all,z=err_inc_max] {Data/err_max.tsv};

\addplot3[matlab1,thick,forget plot] table [x=Nspan_1,y=Nch_1,z=err_max_1] {Data/err_end_max.tsv};
\addplot3[matlab1,thick,forget plot] table [x=Nspan_10,y=Nch_1,z=err_max_10] {Data/err_end_max.tsv};

\addplot3[matlab2,thick,forget plot] table [x=Nspan_1,y=Nch_1,z=err_integral_max_1] {Data/err_end_max.tsv};
\addplot3[matlab2,thick,forget plot] table [x=Nspan_10,y=Nch_1,z=err_integral_max_10] {Data/err_end_max.tsv};

\addplot3[matlab3,thick,forget plot] table [x=Nspan_1,y=Nch_1,z=err_inc_max_1] {Data/err_end_max.tsv};
\addplot3[matlab3,thick,forget plot] table [x=Nspan_10,y=Nch_1,z=err_inc_max_10] {Data/err_end_max.tsv};

\addlegendentry{CFM};
\addlegendimage{surf,matlab1!50,opacity=0.5,faceted color=matlab1!80,line width=0.3pt};
\addlegendentry{Integral};
\addlegendimage{surf,matlab2!50,opacity=0.5,faceted color=matlab2!80,line width=0.3pt};
\addlegendentry{Incoherent};
\addlegendimage{surf,matlab3!50,opacity=0.5,faceted color=matlab3!80,line width=0.3pt};

\end{axis}
\end{tikzpicture}
\caption{Maximum absolute error of the proposed CFM, integral, and incoherent CFM compared to SSFM as a function of different number of spans and optical bandwidth.}
\label{fig:err_max}
\end{figure}
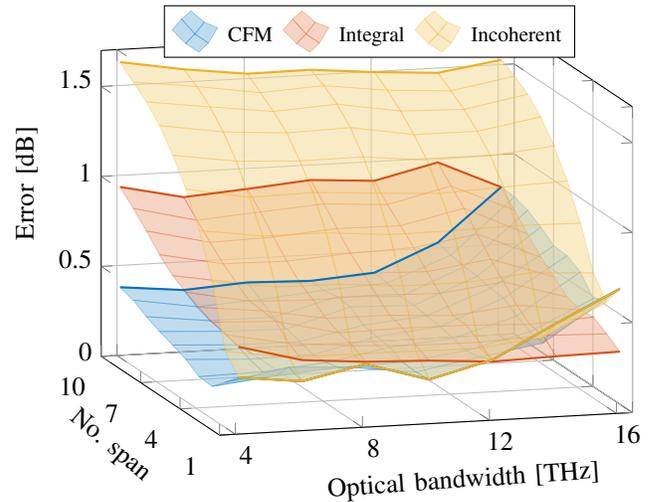

  %
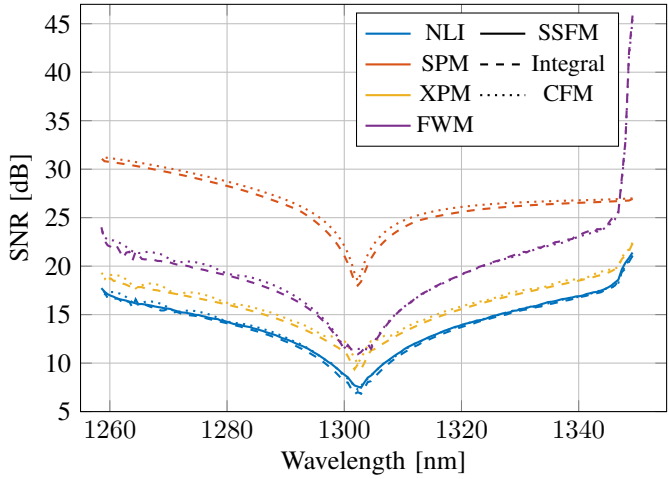
\begin{figure}

\begin{tikzpicture}
    \begin{axis}
    [
    legend columns=2,
    width=0.52\textwidth-1.63326pt,height=7cm,
    grid=both,
    legend style={fill opacity=1, draw opacity=1, text opacity=1, at={(0.47,0.98)}, anchor=north west, draw=black, font=\small},
    xlabel={Wavelength~[nm]},
    ylabel={SNR~[dB]},
    ylabel near ticks,
    xlabel near ticks,
    ylabel shift = -2 pt,
    xlabel shift = -2 pt,
    ytick distance=1,
    ymin=5,
    ymax=47,
    ytick={5,10,15,20,25,30,35,40,45},
    yticklabels={5,10,15,20,25,30,35,40,45},
    xmin=1255,
    xmax=1355,
    clip marker paths=true,
    tick label style={
        /pgf/number format/.cd,
        fixed,
        precision=3,
        use comma=false,
        1000 sep={},
        set decimal separator={.}
    },
    scaled ticks=false,
    axis x line*=box,
    ]

        \addplot[matlab1,thick,solid,mark=none,forget plot] table[x=wvec,y=SNR_SSFM] {Data/snr_10span_-2.tsv};  
        \addplot[matlab1,thick,dashed,mark=none,forget plot] table[x=wvec,y=SNR_NLI_integral] {Data/snr_10span_-2.tsv};
        \addplot[matlab1,thick,dotted,mark=none,forget plot] table[x=wvec,y=SNR_NLI] {Data/snr_10span_-2.tsv};

        \addplot[matlab2,thick,dashed,mark=none,forget plot] table[x=wvec,y=SNR_SPM_integral] {Data/snr_10span_-2.tsv};
        \addplot[matlab2,thick,dotted,mark=none,forget plot] table[x=wvec,y=SNR_SPM] {Data/snr_10span_-2.tsv};

        \addplot[matlab3,thick,dashed,mark=none,forget plot] table[x=wvec,y=SNR_XPM_integral] {Data/snr_10span_-2.tsv};
        \addplot[matlab3,thick,dotted,mark=none,forget plot] table[x=wvec,y=SNR_XPM] {Data/snr_10span_-2.tsv};

        \addplot[matlab4,thick,dashed,mark=none,forget plot] table[x=wvec,y=SNR_FWM_integral] {Data/snr_10span_-2.tsv};
        \addplot[matlab4,thick,dotted,mark=none,forget plot] table[x=wvec,y=SNR_FWM] {Data/snr_10span_-2.tsv};

        \addlegendimage{matlab1,thick,solid}
        \addlegendentry{NLI}
        \addlegendimage{black,thick,solid}
        \addlegendentry{SSFM}
        \addlegendimage{matlab2,thick,solid}
        \addlegendentry{SPM}
        \addlegendimage{black,thick,dashed}
        \addlegendentry{Integral}
        \addlegendimage{matlab3,thick,solid}
        \addlegendentry{XPM}
        \addlegendimage{black,thick,dotted}
        \addlegendentry{CFM}
        \addlegendimage{matlab4,thick,solid}
        \addlegendentry{FWM}
         
    \end{axis}

\end{tikzpicture}

\caption{SPM, XPM and FWM contributions in terms of SNR as a function of wavelength for launch power of -2~dBm per channel in a 10-span transmission.}
\label{fig:snr_10span_-2}

\end{figure} 
We further investigated the accuracy of the proposed CFM with a varying number of spans and transmission bandwidth. In addition to the CFM, the integral model and the incoherent CFM are also compared with SSFM simulations. Figs.~\ref{fig:err} and \ref{fig:err_max} show the mean and maximum absolute error in the $\SNR_{\rm NLI}$, respectively. For the CFM, the mean error is less than 0.22~dB for all scenarios and has a maximum error at 16.1~THz bandwidth and 10 spans. The integral model has a larger mean error at a smaller bandwidth because channels located near the zero-dispersion region present higher error. This error, however, is not reflected in the CFM as it was cancelled by the underestimation of NLI. Note that, this error in the integral model is reduced with increasing bandwidth because channels located further from the zero-dispersion wavelength have a better agreement with SSFM simulations. The larger mean error for the incoherent CFM is mainly caused by the large $\SNR_{\rm NLI}$ error near the zero-dispersion wavelength (see Fig.~\ref{fig:snr_10span_spm_xpm}). CFM with incoherent assumptions shows the larger maximum error of 1.63~dB for 4.1~THz bandwidth at 10 spans. It is reduced to 0.82~dB by using the new CFM given by Eqs.~\eqref{eq:SPM_GN_closed_multispan}~and~\eqref{eq:XPM_GN_closed_multispan}. The CFM error is lower than  that of the integral model because the latter overestimates NLI and CFM underestimates it, which provides less deviation from the results of the SSFM simulations.

\section{Conclusion}
\label{sec:Conclusion}
In this work, a closed-form expression valid for ultra-wideband (UWB) O band nonlinear signal transmission was derived. This was achieved by incorporating four-wave mixing (FWM) contributions into the nonlinear interference (NLI) noise model and introducing a new derivation of the coherent factor for both self-phase modulation (SPM) and cross-phase modulation (XPM). These enhancements are essential for ensuring the accuracy of the formula across any number of spans and dispersion values.

The proposed expression was evaluated for transmission distances ranging from 80 to 800~km and optical bandwidths between 4.1 and 16.1~THz. Its accuracy was benchmarked against both the integral model and split-step Fourier method (SSFM) simulations, yielding a mean absolute error in the NLI signal-to-noise ratio (SNR) below 0.22~dB across all scenarios.

This model enables accurate and rapid estimation of NLI SNR at any distance and optical bandwidth, making it a practical tool for transmission performance evaluation and system optimisation tasks - including optimum launch power, channel symbol rate and spacing, maximum reach, etc. With the proposed formula, NLI can be computed in microseconds on state-of-the-art processors.

\section*{Acknowledgement}
For the purpose of open access, the author has applied a Creative Commons Attribution (CC BY) licence to any Author Accepted Manuscript version arising.

\appendices

\section{Derivation of the link function}
\label{appA:link_function}

This section describes the derivation of Eq.~\eqref{eq:link_function_closed_set3}. Using Eq.~\eqref{eq:FWM_condition}, the link function $\mu (f_1+f_j, f_2 + f_k, f_i)$ in Eq.~\eqref{eq:link_function_integral} can be written as 
\begin{equation}
\begin{split}
&\mu\left(f_1+f_j, f_2 + f_k, f_i\right)=\\
&\left| \int_0^L d\zeta \ \sqrt{\frac{\rho(\zeta,f_1+f_j) \rho(\zeta,f_2 + f_k) \rho(\zeta,f_1 + f_2 + f_m)}{\rho(\zeta,f_i)}} \right. \\
&\left. \times \vphantom{\sqrt{\frac{\rho(\zeta,f_1+f_j) \rho(\zeta,f_2 + f_k) \rho(\zeta,f_1 + f_2 + f_m)}{\rho(\zeta,f_i)}}} e^{j\phi\left(f_1+f_j,f_2 + f_k,f_i\right)\zeta}\right|^2.
\label{appA:eq:link_function_FWM_integral}
\end{split}
\end{equation}
Assuming that the power evolution is constant over the bandwidth of each channel, $f_1 = 0$ and $f_2 = 0$ inside the function $\rho$, the square root of the signal profile for the channel $i$, $\sqrt{\rho(z,f_i)}$, is given by Eq.~\eqref{eq:signal_profile_closed_taylor}, which can be rewritten as
\begin{equation}
\begin{split}
&\sqrt{\rho(z,f_i)} \approx T_i\left(1 - \frac{\tilde{T}_i}{T_i} e^{-\tilde{\alpha}_i \zeta}\right)e^{-\frac{\alpha_i}{2} \zeta} =\\
&T_i\sum_{\substack{l_i \in \{0,1\}}} \left( \frac{-\tilde{T}_i}{T_i} \right)^{l_i} e^{-l_i\tilde{\alpha}_i \zeta}  e^{-\frac{\alpha_i}{2} \zeta},
\label{appA:eq:signal_profile_closed_taylor}
\end{split}
\end{equation}
where the identity in Eq.~\eqref{appE:multinomial_theorem} was used.
For the channels $j$, $k$, and $m$, $\sqrt{\rho(z,f_j)}$, $\sqrt{\rho(z,f_k)}$, and $\sqrt{\rho(z,f_m)}$ are obtained by replacing the index $i$ in Eq.~\eqref{appA:eq:signal_profile_closed_taylor} by $j$, $k$, and $m$. Thus, by using Eq.~\eqref{appA:eq:signal_profile_closed_taylor} and Eq.~\eqref{appA:eq:link_function_FWM_integral} , it is possible to write 
\begin{equation}
\begin{split}
&\mu\left(f_1+f_j, f_2 + f_k, f_i\right)=\left| \int_0^L d\zeta \ \frac{T_jT_kT_m}{T_i} \right. \\
&\left. \times \frac{\left(1 - \frac{\tilde{T}_j}{T_j} e^{-\tilde{\alpha}_j \zeta}\right) \left(1 - \frac{\tilde{T}_k}{T_k} e^{-\tilde{\alpha}_k \zeta}\right) \left(1 - \frac{\tilde{T}_m}{T_m} e^{-\tilde{\alpha}_m \zeta}\right) }{\left(1 - \frac{\tilde{T}_i}{T_i} e^{-\tilde{\alpha}_i \zeta}\right)} \right. \\
&\left. \times e^{\left(\frac{\alpha_j + \alpha_k + \alpha_m - \alpha_i}{2}\right)\zeta}e^{j\phi\zeta}\right|^2,
\label{appA:eq:link_function_FWM_integral1}
\end{split}
\end{equation}
where the frequency dependence of $\phi$ was omitted to simplify notation.

For the different frequency combinations in the set $\Omega_1$, i.e, for the channels $f_j$, $f_k$ and $f_m$ = $f_i$, $m$ = $i$, Eq.~\eqref{appA:eq:link_function_FWM_integral1}, reduces to
\begin{equation}
\begin{split}
&\mu\left(f_1+f_j, f_2 + f_k, f_i\right)= \\ 
&\left| T_jT_k \sum_{\substack{l_j, l_k \in \{0,1\}}} \left( \frac{-\tilde{T}_j}{T_j} \right)^{l_j} \left( \frac{-\tilde{T}_k}{T_k} \right)^{l_k} \right. \\ 
&\left. \times \int_0^L d\zeta \ \ e^{-\left(l_j\tilde{\alpha}_j+ l_k\tilde{\alpha}_k + \frac{\alpha_j + \alpha_k}{2}\right)\zeta + j \phi \zeta}\right|^2,
\label{appA:eq:link_function_FWM_integral2}
\end{split}
\end{equation}
where the solution of this integral is given by 
\begin{equation}
\begin{split}
&\mu\left(f_1+f_j, f_2 + f_k, f_i\right)= \left| T_j T_k \sum_{\substack{l_j, l_k \in \{0, 1\}}} \left( \frac{-\tilde{T}_j}{T_j} \right)^{l_j} \right. \\
&\left. \times \left( \frac{-\tilde{T}_k}{T_k} \right)^{l_k} \frac{1 - e^{-\left(l_j\tilde{\alpha}_j+ l_k\tilde{\alpha}_k + \frac{\alpha_j + \alpha_k}{2}\right)L + j\phi L}}{-\left(l_j\tilde{\alpha}_j+ l_k\tilde{\alpha}_k + \frac{\alpha_j + \alpha_k}{2}\right) + j\phi}
\right|^2.
\label{appA:eq:link_function_closed_set1}
\end{split}
\end{equation}

For frequency combinations in the set $\Omega_2$, Eq.~\eqref{appA:eq:link_function_FWM_integral1} has to be used. To be able to integrate this equation analytically, $\tilde{T}_i = 0$ is assumed, yielding $T_i = 1 + \tilde{T}_i = 1$. This assumption is equivalent to neglecting part of the ISRS effect for the channel $i$ only, which is jointly modelled by the fitting coefficients $C_{r,i}$, $\tilde{\alpha}_i$ and $\alpha_i$~\cite{Buglia_JLT_lowloss}. Using this approximation Eq.~\eqref{appA:eq:link_function_FWM_integral1} can be rewritten as
\begin{equation}
\begin{split}
&\mu\left(f_1+f_j, f_2 + f_k, f_i\right)= \\ 
&\left| T_jT_kT_m  \sum_{\substack{l_j, l_k, l_m \in \{0,1\}}} \left( \frac{-\tilde{T}_j}{T_j} \right)^{l_j} \left( \frac{-\tilde{T}_k}{T_k} \right)^{l_k} \left( \frac{-\tilde{T}_m}{T_m} \right)^{l_m} \right. \\
&\left. \times \int_0^L d\zeta \ \ e^{-\left(l_j\tilde{\alpha}_j+ l_k\tilde{\alpha}_k + l_m\tilde{\alpha}_m + \frac{\alpha_j + \alpha_k + \alpha_m - \alpha_i}{2}\right)\zeta + j \phi \zeta}\right|^2,
\label{appA:eq:link_function_FWM_integral3}
\end{split}
\end{equation}
where the solution of this integral is given by 
\begin{equation}
\begin{split}
&\mu\left(f_1+f_j, f_2 + f_k, f_i\right)= \\
&\left| T_j T_k T_m \sum_{\substack{l_j, l_k, l_m \in \{0, 1\}}} \left( \frac{-\tilde{T}_j}{T_j} \right)^{l_j} \left( \frac{-\tilde{T}_k}{T_k} \right)^{l_k} \left( \frac{-\tilde{T}_m}{T_m} \right)^{l_m} \right. \\
&\left. \times \frac{1 - e^{-\left(l_j\tilde{\alpha}_j+ l_k\tilde{\alpha}_k + l_m\tilde{\alpha}_m + \frac{\alpha_j + \alpha_k + \alpha_m - \alpha_i}{2}\right)L + j\phi L}}{-\left(l_j\tilde{\alpha}_j+ l_k\tilde{\alpha}_k + l_m\tilde{\alpha}_m + \frac{\alpha_j + \alpha_k + \alpha_m - \alpha_i}{2}\right) + j\phi}
\right|^2.
\label{eq:link_function_closed_set2}
\end{split}
\end{equation}
In order to solve Eq.~\eqref{eq:GN_integral_gaussian} in closed-form, the approach in~\cite{zefreh2021gnmodelclosedformformulasupporting}, which approximate the fraction with exponential terms, is used as follows
\begin{equation}
\frac{1 - e^{-\alpha_{\ell, i}L + j\phi L}}{-\alpha_{\ell, i} + j\phi} \approx \frac{\kappa_{\ell, i}}{-\tilde{\alpha}_{\ell, i} + j\phi},
\label{appA:eq:link_function_FWM_fraction}
\end{equation}
where $\tilde{\alpha}_{\ell, i}$ and $\kappa_{\ell, i}$ are chosen such that the first-order Taylor approximations of both the left and the right side of Eq.~\eqref{appA:eq:link_function_FWM_fraction} around the variable $\phi = 0$ become equal. This yields Eqs.~\eqref{eq:alpha_tilde}~and~\eqref{eq:kappa} with notation simplification using Eqs.~\eqref{eq:link_simplify_l},~\eqref{eq:link_simplify_d},~\eqref{eq:link_simplify_a}~and~\eqref{eq:link_simplify_T}. Replacing the approximation in Eq.~\eqref{appA:eq:link_function_FWM_fraction} into Eq.~\eqref{eq:link_function_closed_set2}, yields
\begin{equation}
\begin{split}
&\mu\left(f_1+f_j, f_2 + f_k, f_i\right)= \\
&\left(\sum_{\substack{\ell}} \mathcal{T}_{\ell} \frac{\kappa_{\ell, i}}{-\tilde{\alpha}_{\ell, i} + j\phi}\right) \times \left(\sum_{\substack{\ell'}} \mathcal{T}_{\ell'} \frac{\kappa_{\ell', i}}{-\tilde{\alpha}_{\ell', i} - j\phi}\right).
\label{appA:eq:link_function_closed_set3}
\end{split}
\end{equation}
Finally, performing the multiplication in Eq.~\eqref{appA:eq:link_function_closed_set3} yields Eq.~\eqref{eq:link_function_closed_set3}, concluding the proof.

\section{Derivation of the FWM contribution.}
\label{appB:FWM}

This section presents the derivation of Eq.~\eqref{eq:FWM_GN_closed}. The phase mismatch term $\phi\left(f_1+f_j,f_2 + f_k,f_i\right)$ is firstly approximated. Let $\Delta f_j = f_j - f_i$ and $\Delta f_k = f_k - f_i$ be the frequency separation between channels $j$ and $i$, and between channels $k$ and $i$ respectively. $\phi\left(f_1+f_j,f_2 + f_k,f_i\right)$ can thus be written as
\begin{equation}
\begin{split}
&\phi(f_1+f_j,f_2+f_k,f_i) = -4\pi^2\left(f_1+\Delta f_j\right)\left(f_2+\Delta f_k\right) \\
&\times \left[\beta_2+\pi\beta_3\left(f_1+f_j+f_2+f_k\right) + \frac{2\pi^2}{3} \beta_4\left(\left(f_1 + \Delta f_j\right)^2 \right. \right. \\
&\left. \left. + \frac{3}{2}\left(f_1 + \Delta f_j\right)\left(f_2+\Delta f_k\right) + 3\left(f_1 + \Delta f_j\right)f_i \right. \right. \\ 
&\left. \vphantom{\frac{2\pi^2}{3}} \left. + \left(f_2 + \Delta f_k\right)^2 + 3\left(f_2 + \Delta f_k\right)f_i + 3f_i^2\right)\right].
\label{appB:mismatch_term_approx1}
\end{split}
\end{equation}
By considering a first-order two-dimensional Taylor approximation around $\left(f_1,f_2\right) = \left(0,0\right)$, it can be approximated as $\phi(f_1+f_j,f_2+f_k,f_i) \approx \phi_0+\phi_1 f_1+\phi_2 f_2$, where
\begin{equation}
\begin{split}
&\phi_0 = -4\pi^2\Delta f_j\Delta f_k\left[\beta_2 + \pi\beta_3\left(f_j+f_k\right) + \frac{2\pi^2}{3}\beta_4 Q_0\right], \\
&\phi_1 = -4\pi^2\Delta f_k\left[\beta_2 + \pi\beta_3\left(f_j+f_k+\Delta f_j\right) \right. \\
&\left. + \frac{2\pi^2}{3}\beta_4\left(Q_0+\Delta f_j Q_1\right)\right], \\
&\phi_2 = -4\pi^2\Delta f_j\left[\beta_2 + \pi\beta_3\left(f_j+f_k+\Delta f_k\right) \right. \\
&\left. + \frac{2\pi^2}{3}\beta_4\left(Q_0+\Delta f_k Q_2\right)\right], \\
\label{appB:mismatch_term_approx2}
\end{split}
\end{equation}
and
\begin{equation}
\begin{split}
&Q_0 = \Delta f_j^2 + \frac{3}{2}\Delta f_j\Delta f_k + 3\Delta f_j f_i + \Delta f_k^2 + 3\Delta f_k f_i + 3f_i^2, \\
&Q_1 = 2\Delta f_j + \frac{3}{2}\Delta f_k + 3f_i, \\
&Q_2 = 2\Delta f_k + \frac{3}{2}\Delta f_j + 3f_i. \\
\label{appB:mismatch_term_approx3}
\end{split}
\end{equation}
Thus, the FWM contributions to the NLI given by Eq.~\eqref{eq:FWM_GN_integral} can written as
\begin{equation}
\begin{split}
&\eta_{\text{FWM}}^{(j,k,m)}(f_i) = \frac{16}{27} \gamma^2 \frac{ B_i}{P_i^3} \frac{P_j P_k P_m}{B_j B_k B_m} \\
&\times\sum_{\substack{\ell, \ell'}} \mathcal{T}_{\ell}\mathcal{T}_{\ell'}\kappa_{\ell, i}\kappa_{\ell', i} \int_{\frac{-B_j}{2}}^{\frac{B_j}{2}} df_1 \int_{\frac{-B_k}{2}}^{\frac{B_k}{2}} df_2\  \Pi \left(\frac{f_1 + f_2}{B_m} \right) \\
&\times\frac{\tilde{\alpha}_{\ell, i}\tilde{\alpha}_{\ell', i}+\left(\phi_0+\phi_1 f_1+\phi_2 f_2\right)^2}{\left(\tilde{\alpha}_{\ell, i}^2+\left(\phi_0+\phi_1 f_1+\phi_2 f_2\right)^2\right)\left(\tilde{\alpha}_{\ell', i}^2+\left(\phi_0+\phi_1 f_1+\phi_2 f_2\right)^2\right)}.
\label{appB:eq:FWM_GN_integral}
\end{split}
\end{equation}
By neglecting the term $\Pi \left(\frac{f_1 + f_2}{B_m} \right)$, i.e., by assuming that the integration domain can be approximated as a rectangle, the first of the two integrals in Eq.~\eqref{appB:eq:FWM_GN_integral} can be solved by using identity \eqref{appE:atan}, yielding
\begin{equation}
\begin{split}
&\int_{\frac{-B_k}{2}}^{\frac{B_k}{2}} df_2 \frac{1}{\phi_2\left(\tilde{\alpha}_{\ell, i}+\tilde{\alpha}_{\ell', i}\right)} \left[ \atan\left(\frac{2\phi_0+\phi_1 B_j}{2\tilde{\alpha}_{\ell, i}}+\frac{\phi_2}{\tilde{\alpha}_{\ell, i}}f_2\right) \right. \\
&+ \atan\left(\frac{2\phi_0+\phi_1 B_j}{2\tilde{\alpha}_{\ell', i}}+\frac{\phi_2}{\tilde{\alpha}_{\ell', i}}f_2\right) \\
&- \atan\left(\frac{2\phi_0-\phi_1 B_j}{2\tilde{\alpha}_{\ell, i}}+\frac{\phi_2}{\tilde{\alpha}_{\ell, i}}f_2\right) \\
&\left. - \atan\left(\frac{2\phi_0-\phi_1 B_j}{2\tilde{\alpha}_{\ell', i}}+\frac{\phi_2}{\tilde{\alpha}_{\ell', i}}f_2\right) \right].
\label{appB:eq:FWM_GN_integral2}
\end{split}
\end{equation}
Each integral containing $\atan$ term in Eq.~\eqref{appB:eq:FWM_GN_integral2} can be solved by using the identity in Eq.~\eqref{appE:atan_integral}, yielding
\begin{equation}
\begin{split}
&\eta_{\text{FWM}}^{(j,k,m)}(f_i) = \sum_{\substack{\ell, \ell'}} \frac{\mathcal{T}_{\ell}\mathcal{T}_{\ell'}\kappa_{\ell, i}\kappa_{\ell', i}}{\phi_1\phi_2\left(\tilde{\alpha}_{\ell, i}+\tilde{\alpha}_{\ell', i}\right)} \\
&\times \left[ \tilde{\alpha}_{\ell, i}\left(F\left(u_{+}\right)-F\left(u_{-}\right)-F\left(u'_{+}\right)+F\left(u'_{-}\right)\right) \right. \\
&\left. + \tilde{\alpha}_{\ell', i}\left(F\left(v_{+}\right)-F\left(v_{-}\right)-F\left(v'_{+}\right)+F\left(v'_{-}\right)\right) \right],
\label{appB:eq:FWM_GN_closed}
\end{split}
\end{equation}
where $F\left(x\right) = x\atan\left(x\right)-\frac{1}{2}\ln\left(1+x^2\right)$ and 
\begin{equation}
\begin{aligned} 
u_{\pm} &= \frac{2\phi_0+\phi_1 B_j\pm\phi_2 B_k}{2\tilde{\alpha}_{\ell, i}},    & v_{\pm} &= \frac{2\phi_0-\phi_1 B_j\pm\phi_2 B_k}{2\tilde{\alpha}_{\ell, i}}, \\ 
u'_{\pm} &= \frac{2\phi_0+\phi_1 B_j\pm\phi_2 B_k}{2\tilde{\alpha}_{\ell', i}}, & v'_{\pm} &= \frac{2\phi_0-\phi_1 B_j\pm\phi_2 B_k}{2\tilde{\alpha}_{\ell', i}}. 
\end{aligned}
\end{equation}
By inserting the pre-factor $\frac{16}{27} \gamma^2 \frac{ B_i}{P_i^3} \frac{P_j P_k P_m}{B_j B_k B_m}$ in Eq.~\eqref{appB:eq:FWM_GN_closed}, Eq.~\eqref{eq:FWM_GN_closed} is obtained, the proof is concluded.

\section{Derivation of the multi-span SPM coherent contribution.}
\label{appC:SPM_multispan}

This section presents the derivation of Eq.~\eqref{eq:SPM_GN_closed_multispan}. The coherent contribution of the nonlinear coefficient requires re-deriving the integral as follow
\begin{equation}
\begin{split}
&\eta_{\rm SPM,cc}(f_i) = {T'_i}^2 \sum_{\substack{l, l' \in \{0, 1\}}} \left( \frac{-\tilde{T}'_i}{T'_i} \right)^{l+l'} \kappa_{l,i} \kappa_{l',i} \sum_{n=1}^{N_s - 1} 2\left(N_s - n\right) \\
&\times \int_{\frac{-B_i}{2}}^{\frac{B_i}{2}} df_1 \int_{\frac{-B_i}{2}}^{\frac{B_i}{2}} df_2\ \frac{\left(\tilde{\alpha}_{l,i} \tilde{\alpha}_{l',i} + \phi_i^2 f_1^2 f_2^2\right) \cos\left(n \phi_i f_1 f_2 L\right)}{\left(\tilde{\alpha}_{l,i}^2 + \phi_i^2 f_1^2 f_2^2\right)\left(\tilde{\alpha}_{l',i}^2 + \phi_i^2 f_1^2 f_2^2\right)},
\label{appC:eq:SPM_GN_integral_multispan}
\end{split}
\end{equation}
where $\phi_i$ can be found in \cite[Eq.~(27)]{Min_JLT} and $\tilde{T}'_i = -\frac{P_{\text{tot}}C_{r,i}}{\tilde{\alpha}}f_i$, $T'_i = 1 + \tilde{T}'_i$. The integrand in Eq.~\eqref{appC:eq:SPM_GN_integral_multispan} can be approximated as a damped function given the assumption that $\tilde{\alpha}_{l,i}^2 +  \tilde{\alpha}_{l',i}^2 \approx 2\tilde{\alpha}_{l,i}\tilde{\alpha}_{l',i}$ which holds for every channel $i$, as shown below
\begin{equation}
\frac{\left(\tilde{\alpha}_{l,i} \tilde{\alpha}_{l',i} + \phi_i^2 f_1^2 f_2^2\right) \cos\left(n \phi_i f_1 f_2 L\right)}{\left(\tilde{\alpha}_{l,i}^2 + \phi_i^2 f_1^2 f_2^2\right)\left(\tilde{\alpha}_{l',i}^2 + \phi_i^2 f_1^2 f_2^2\right)} \approx \frac{ \cos\left(n \phi_i f_1 f_2 L\right)}{\tilde{\alpha}_{l,i}\tilde{\alpha}_{l',i} + \phi_i^2 f_1^2 f_2^2}.
\label{appC:eq:SPM_integrand}
\end{equation}
It allows the first of the two integrals to be solved by using identity \eqref{appE:exponential_integral} as
\begin{equation}
\begin{split}
&\eta_{\rm SPM,cc}(f_i) = {T'_i}^2 \sum_{\substack{l, l' \in \{0, 1\}}} \left( \frac{-\tilde{T}'_i}{T'_i} \right)^{l+l'} \frac{\kappa_{l,i} \kappa_{l',i}}{\phi_i\sqrt{\tilde{\alpha}_{l,i} \tilde{\alpha}_{l',i}}} \\
&\times \sum_{n=1}^{N_{\rm s} - 1} 2\left(N_{\rm s} - n\right) \int_{\frac{-B_i}{2}}^{\frac{B_i}{2}} df_2\ \frac{1}{f_2} \left[ \sign(\phi_i f_2) \pi e^{-nL\sqrt{\tilde{\alpha}_{l,i} \tilde{\alpha}_{l',i}}} \vphantom{+ e^{nL\sqrt{\tilde{\alpha}_{l,i} \tilde{\alpha}_{l',i}}} \  \operatorname{Im} \left\{ E_1 \left(nL\left[\sqrt{\tilde{\alpha}_{l,i} \tilde{\alpha}_{l',i}} - j\phi_i f_2\frac{B_i}{2} \right]\right) \right\}} \right. \\
&\left. + e^{nL\sqrt{\tilde{\alpha}_{l,i} \tilde{\alpha}_{l',i}}} \  \operatorname{Im} \left\{ E_1 \left(nL\left[\sqrt{\tilde{\alpha}_{l,i} \tilde{\alpha}_{l',i}} - j\phi_i f_2\frac{B_i}{2} \right]\right) \right\} \right. \\
&\left. - e^{-nL\sqrt{\tilde{\alpha}_{l,i}\tilde{\alpha}_{l',i}}} \  \operatorname{Im} \left\{ E_1 \left(-nL\left[\sqrt{\tilde{\alpha}_{l,i}\tilde{\alpha}_{l',i}} + j\phi_i f_2\frac{B_i}{2} \right]\right) \right\} \right],
\label{appC:eq:SPM_GN_integral_multispan2}
\end{split}
\end{equation}
where $E_1\left(\cdot\right)$ is the exponential integral function and $\operatorname{Im}\left\{\cdot\right\}$ takes the imaginary part of the complex value. Assuming $\tilde{\alpha}_{l,i}L \gg 1$ and $\tilde{\alpha}_{l',i}L \gg 1$, the first term in the integrand that comes from the branch-cut of $E_1$ is suppressed exponentially and can be neglected in the subsequent derivation. The magnitude of the argument of $E_1$ is much greater than one. We can then get the following approximation given the first-order asymptotic expansion of $E_1\left(x\right) \approx \frac{e^{-x}}{x}$
\begin{equation}
\begin{split}
&\eta_{\rm SPM,cc}(f_i) \approx {T'_i}^2 \sum_{\substack{l, l' \in \{0, 1\}}} \left( \frac{-\tilde{T}'_i}{T'_i} \right)^{l+l'} \frac{\kappa_{l,i} \kappa_{l',i}}{\phi_i\sqrt{\tilde{\alpha}_{l,i} \tilde{\alpha}_{l',i}}} \\
&\times \sum_{n=1}^{N_{\rm s} - 1} 2\left(N_{\rm s} - n\right) \int_{\frac{-B_i}{2}}^{\frac{B_i}{2}} df_2\ \frac{2\sqrt{\tilde{\alpha}_{l,i} \tilde{\alpha}_{l',i}}\sin\left(n\phi_i f_2 L\frac{B_i}{2}\right)}{nL\left(\tilde{\alpha}_{l,i}\tilde{\alpha}_{l',i}+\phi_i^2 f_2^2\frac{B_i^2}{4}\right)f_2} .
\label{appC:eq:SPM_GN_integral_multispan3}
\end{split}
\end{equation}
The integrand in Eq.~\eqref{appC:eq:SPM_GN_integral_multispan3} can be further simplified by applying partial fraction, as follows
\begin{equation}
\begin{split}
&\eta_{\rm SPM,cc}(f_i) \approx {T'_i}^2 \sum_{\substack{l, l' \in \{0, 1\}}} \left( \frac{-\tilde{T}'_i}{T'_i} \right)^{l+l'} \frac{\kappa_{l,i} \kappa_{l',i}}{\phi_i L} \\
&\times \sum_{n=1}^{N_{\rm s} - 1} \frac{4\left(N_{\rm s} - n\right)}{n} \left[ \frac{1}{\tilde{\alpha}_{l,i}\tilde{\alpha}_{l',i}} \int_{\frac{-B_i}{2}}^{\frac{B_i}{2}} df_2\ \frac{\sin\left(n\phi_i f_2 L\frac{B_i}{2}\right)}{f_2} \right. \\
&- \left. \frac{\phi_i^2\frac{B_i^2}{4}}{\tilde{\alpha}_{l,i}\tilde{\alpha}_{l',i}} \int_{\frac{-B_i}{2}}^{\frac{B_i}{2}} df_2\ \frac{\sin\left(n\phi_i f_2 L\frac{B_i}{2}\right)f_2}{\left(\tilde{\alpha}_{l,i}\tilde{\alpha}_{l',i}+\phi_i^2 f_2^2\frac{B_i^2}{4}\right)} \right].
\label{appC:eq:SPM_GN_integral_multispan4}
\end{split}
\end{equation}
The second integral can be written as
\begin{equation}
\begin{split}
&\frac{\phi_i^2\frac{B_i^2}{4}}{\tilde{\alpha}_{l,i}\tilde{\alpha}_{l',i}} \int_{\frac{-B_i}{2}}^{\frac{B_i}{2}} df_2\ \frac{\sin\left(n\phi_i f_2 L\frac{B_i}{2}\right)f_2}{\left(\tilde{\alpha}_{l,i}\tilde{\alpha}_{l',i}+\phi_i^2 f_2^2\frac{B_i^2}{4}\right)} \\
&= \frac{1}{\tilde{\alpha}_{l,i}\tilde{\alpha}_{l',i}} \int_{\frac{-B_i}{2}}^{\frac{B_i}{2}} df_2\ \frac{\sin\left(n\phi_i f_2 L\frac{B_i}{2}\right)}{\frac{\tilde{\alpha}_{l,i}\tilde{\alpha}_{l',i}}{\phi_i^2 f_2\frac{B_i^2}{4}}+f_2}.
\label{appC:eq:SPM_GN_integral_multispan5}
\end{split}
\end{equation}
As $\tilde{\alpha}_{l,i}\tilde{\alpha}_{l',i} \gg \phi_i^2 f_2\frac{B_i^2}{4}$, the first integral in Eq.~\eqref{appC:eq:SPM_GN_integral_multispan4} is dominant which leads to an solution using the identity~\eqref{appE:sine_integral}
\begin{equation}
\begin{split}
&\eta_{\rm SPM,cc}(f_i) = {T'_i}^2 \sum_{\substack{l, l' \in \{0, 1\}}} \left( \frac{-\tilde{T}'_i}{T'_i} \right)^{l+l'} \frac{\kappa_{l,i} \kappa_{l',i}}{\phi_i L\tilde{\alpha}_{l,i}\tilde{\alpha}_{l',i}} \\
&\times \sum_{n=1}^{N_{\rm s} - 1} \frac{8\left(N_{\rm s} - n\right)}{n} \Si\left(n\phi_i L\frac{B_i^2}{4}\right),
\label{appC:eq:SPM_GN_closed_multispan6}
\end{split}
\end{equation}
where $\Si\left(\cdot\right)$ is the sine integral function. To avoid such special function, we approximate $\Si\left(x\right) \approx \atan\left(x\right)$ for small argument and
\begin{equation}
\begin{split}
&\eta_{\rm SPM,cc}(f_i) \approx {T'_i}^2 \sum_{\substack{l, l' \in \{0, 1\}}} \left( \frac{-\tilde{T}'_i}{T'_i} \right)^{l+l'} \frac{\kappa_{l,i} \kappa_{l',i}}{\phi_i L\tilde{\alpha}_{l,i}\tilde{\alpha}_{l',i}} \\
&\times \sum_{n=1}^{N_{\rm s} - 1} \frac{8\left(N_{\rm s} - n\right)}{n} \atan\left(n\phi_i L\frac{B_i^2}{4}\right).
\label{appC:eq:SPM_GN_closed_multispan}
\end{split}
\end{equation}
By inserting the pre-factor $\frac{16}{27}\frac{\gamma^2}{B_i^2}$ in Eq.~\eqref{appC:eq:SPM_GN_closed_multispan}, Eq.~\eqref{eq:SPM_GN_closed_multispan} is obtained, concluding the proof.

\section{Derivation of the multi-span XPM coherent contribution.}
\label{appD:XPM_multispan}

The nonlinear coefficient XPM coherent contribution with similar approximation of the integrand in Eq.~\eqref{appC:eq:SPM_integrand} can be written as
\begin{equation}
\begin{split}
&\eta_{\rm XPM,cc}(f_i) = {T'_i}^2 \sum_{\substack{l, l' \in \{0, 1\}}} \left( \frac{-\tilde{T}'_k}{T'_k} \right)^{l+l'} \kappa_{l,k} \kappa_{l',k} \\
&\times \sum_{n=1}^{N_s - 1} 2\left(N_s - n\right) 2 B_k \int_{0}^{\frac{B_i}{2}} df_1\ \frac{\cos\left(n \phi_{i,k} f_1 L\right)}{\tilde{\alpha}_{l,k}^2\tilde{\alpha}_{l',k}^2 + \phi_{i,k}^2 f_1^2},
\label{appD:eq:XPM_GN_integral_multispan2}
\end{split}
\end{equation}
where $\phi_{i,k}$ can be found in \cite[Eq.~(26)]{Min_JLT}. It has an exact solution using identity \eqref{appE:exponential_integral} as
\begin{equation}
\begin{split}
&\eta_{\rm XPM,cc}(f_i) = {T'_i}^2 \sum_{\substack{l, l' \in \{0, 1\}}} \left( \frac{-\tilde{T}'_k}{T'_k} \right)^{l+l'} \frac{B_k\kappa_{l,k} \kappa_{l',k}}{\phi_{i,k}\sqrt{\tilde{\alpha}_{l,k} \tilde{\alpha}_{l',k}}} \\
&\times \sum_{n=1}^{N_{\rm s} - 1} 2\left(N_{\rm s} - n\right) \left[ \sign(\phi_{i,k}) \pi e^{-nL\sqrt{\tilde{\alpha}_{l,k} \tilde{\alpha}_{l',k}}} \vphantom{+ e^{nL\sqrt{\tilde{\alpha}_{l,k} \tilde{\alpha}_{l',k}}} \  \operatorname{Im} \left\{ E_1 \left(nL\left[\sqrt{\tilde{\alpha}_{l,k} \tilde{\alpha}_{l',k}} - j\phi_{i,k}\frac{B_i}{2} \right]\right) \right\}} \right. \\
&\left. + e^{nL\sqrt{\tilde{\alpha}_{l,k} \tilde{\alpha}_{l',k}}} \  \operatorname{Im} \left\{ E_1 \left(nL\left[\sqrt{\tilde{\alpha}_{l,k} \tilde{\alpha}_{l',k}} - j\phi_{i,k}\frac{B_i}{2} \right]\right) \right\} \right. \\
&\left. - e^{-nL\sqrt{\tilde{\alpha}_{l,k}\tilde{\alpha}_{l',k}}} \  \operatorname{Im} \left\{ E_1 \left(-nL\left[\sqrt{\tilde{\alpha}_{l,k}\tilde{\alpha}_{l',k}} - j\phi_{i,k}\frac{B_i}{2} \right]\right) \right\} \right].
\label{appD:eq:XPM_GN_closed_multispan2}
\end{split}
\end{equation}
Similarly to Eq.~\eqref{appC:eq:SPM_GN_integral_multispan2}, Eq.~\eqref{appD:eq:XPM_GN_closed_multispan2} can be simplified as
\begin{equation}
\begin{split}
&\eta_{\rm XPM,cc}(f_i) \approx {T'_i}^2 \sum_{\substack{l, l' \in \{0, 1\}}} \left( \frac{-\tilde{T}'_k}{T'_k} \right)^{l+l'} \frac{B_k\kappa_{l,k} \kappa_{l',k}}{\phi_{i,k}\sqrt{\tilde{\alpha}_{l,k} \tilde{\alpha}_{l',k}}} \\
&\times \sum_{n=1}^{N_{\rm s} - 1} 2\left(N_{\rm s} - n\right) \left[\sign(\phi_{i,k}) \pi e^{-nL\sqrt{\tilde{\alpha}_{l,k} \tilde{\alpha}_{l',k}}} \vphantom{\frac{2\sqrt{\tilde{\alpha}_{l,k} \tilde{\alpha}_{l',k}}\sin\left(j n\phi_{i,k} L\frac{B_i}{2}\right)}{nL\left(\tilde{\alpha}_{l,k}\tilde{\alpha}_{l',k}+\phi_{i,k}^2\frac{B_i^2}{4}\right)}} \right. \\
&\left. + \frac{2\sqrt{\tilde{\alpha}_{l,k} \tilde{\alpha}_{l',k}}\sin\left(j n\phi_{i,k} L\frac{B_i}{2}\right)}{nL\left(\tilde{\alpha}_{l,k}\tilde{\alpha}_{l',k}+\phi_{i,k}^2\frac{B_i^2}{4}\right)} \right].
\label{appD:eq:XPM_GN_closed_multispan}
\end{split}
\end{equation}
It seems the above expression cannot be further simplified where the branch-cut term of $E_1$ cannot be neglected to maintain the pointwise accuracy of the expression. As the approximation error of the phase-mismatch from higher order dispersion term will be accumulated in multi-span scenario, it's suggested including it in calculating single-span but not in coherent contribution for multi-span scenarios. By inserting the pre-factor $\frac{32}{27}\frac{\gamma^2}{B_k^2}\left(\frac{P_k}{P_i}\right)^2$ in Eq.~\eqref{appD:eq:XPM_GN_closed_multispan}, Eq.~\eqref{eq:XPM_GN_closed_multispan} is obtained, concluding the proof.

\section{Mathematical Identities}
\label{appE:mathematical_identities}

\begin{equation}
(x + y)^i = \sum_{0 \leq l \leq i} \frac{i!}{l!(i-l)!} x^{l}y^{i-l}.
\label{appE:multinomial_theorem}
\end{equation}

\begin{equation}
\begin{split}
&\int_0^x d\xi\ \frac{ab+c^2\xi^2}{\left(a^2+c^2\xi^2\right)\left(b^2+c^2\xi^2\right)} \\
&= \frac{1}{c\left(a+b\right)}\left[\atan\left(\frac{cx}{a}\right) + \atan\left(\frac{cx}{b}\right)\right].
\label{appE:atan}
\end{split}
\end{equation}

\begin{equation}
\begin{split}
&\int_0^x d\xi\ \frac{\cos\left(\xi\right)}{\left(a^2+\xi^2\right)} = \frac{1}{2a}\left[e^{a}\operatorname{Im}\left\{E_1\left(a-jx\right)\right\} \right. \\
&\left. - e^{-a}\operatorname{Im}\left\{E_1\left(-a-jx\right)\right\} + \sign\left(a\right)\sign\left(x\right)\pi e^{-\left|a\right|}\right].
\label{appE:exponential_integral}
\end{split}
\end{equation}

\begin{equation}
\Si\left(x\right) = \int_{0}^{x}\frac{\sin{\left(\xi\right)}}{\xi}.
\label{appE:sine_integral}
\end{equation}

\begin{equation}
\begin{split}
&\int_{-x}^{x} d\xi\ \atan\left(a+b\xi\right) = \frac{1}{b}\int_{a-bx}^{a+bx} d\xi\ \atan\left(\xi\right) \\
&= \frac{1}{b} \left[\left(a+bx\right)\atan\left(a+bx\right) - \left(a-bx\right)\atan\left(a-bx\right)\right] \\
&- \frac{1}{2}\ln\left(\frac{1+\left(a+bx\right)^2}{1+\left(a-bx\right)^2}\right).
\label{appE:atan_integral}
\end{split}
\end{equation}

\bibliographystyle{IEEEtran}
\bibliography{bibfile}

\end{document}